\documentclass[twocolumn]{aastex631}
\shorttitle{H$\alpha$ emitters revealed by JWST}
\shortauthors{van Mierlo et al.}
\graphicspath{{./}{figures/}}
\usepackage[super]{nth}
\usepackage{amssymb}
\usepackage{multirow}

\newcommand{\highlight}[1]{\textbf{#1}}
\renewcommand{\highlight}[1]{#1}

\begin{document}

\title{A high incidence of dusty H$\alpha$ emitters at $z>3$ Among UltraVISTA dropout galaxies in COSMOS \\revealed by JWST}

\correspondingauthor{Sophie E. van Mierlo}
\email{mierlo@astro.rug.nl}

\author[0000-0001-8289-2863]{Sophie E. van Mierlo}
\affiliation{Kapteyn Astronomical Institute, University of Groningen, PO Box 800, 9700 AV Groningen, The Netherlands}

\author[0000-0001-8183-1460]{Karina I. Caputi }
\affiliation{Kapteyn Astronomical Institute, University of Groningen, PO Box 800, 9700 AV Groningen, The Netherlands}
\affiliation{Cosmic Dawn Center (DAWN), Denmark}

\author[0000-0002-3993-0745]{Matthew L. N. Ashby}
\affiliation{Optical and Infrared Astronomy Division
Center for Astrophysics $|$ Harvard \& Smithsonian
Cambridge, MA 02138 USA}

\author[0000-0002-5588-9156]{Vasily Kokorev}
\affiliation{Kapteyn Astronomical Institute, University of Groningen, PO Box 800, 9700 AV Groningen, The Netherlands}

\author[0000-0001-6066-4624]{Rafael Navarro-Carrera}
\affiliation{Kapteyn Astronomical Institute, University of Groningen, PO Box 800, 9700 AV Groningen, The Netherlands}

\author[0000-0002-5104-8245]{Pierluigi Rinaldi}
\affiliation{Kapteyn Astronomical Institute, University of Groningen, PO Box 800, 9700 AV Groningen, The Netherlands}


\begin{abstract}
We have \highlight{characterized} \highlight{26} Spitzer/IRAC-selected sources from the SMUVS program that are undetected in the UltraVISTA DR5 $H$- and/or $K_\mathrm{s}$-band images, \highlight{covering} 94 arcmin$^2$ of the COSMOS field, which have deep multiwavelength JWST \highlight{photometry}. We analyzed the JWST/NIRCam imaging from the PRIMER survey and ancillary HST data to reveal the properties of these galaxies from spectral energy distribution \highlight{(SED)} fitting. We find that \highlight{the majority} of these galaxies are detected by NIRCam at $\lambda < 2 \, \rm \mu{m}$, with only \highlight{four} remaining as near-infrared dropouts in the deeper JWST images.  Our results indicate that the UltraVISTA dropouts candidates are \highlight{primarily} located at $z>3$ and are characterized by high dust extinctions, with a typical color excess \highlight{$E(B-V) = 0.5 \pm 0.3$} and stellar mass \highlight{$\log(M_*/M_\odot) = 9.5 \pm 1.0$}. Remarkably, \highlight{$\sim 75$}\,\% of these sources show a flux enhancement \highlight{between the observed photometry and modeled continuum SED} that can be attributed to H$\alpha$ emission in the corresponding NIRCam bands. The derived (H$\alpha$+ N[II] + S[II]) rest-frame equivalent widths and H$\alpha$ star formation rates (SFRs) span values \highlight{$\sim 100$--2200\,\AA\ and $\sim 5$--375} $M_\odot\, \mathrm{yr^{-1}}$, respectively. The location of these sources on the SFR-$M_*$ plane indicates that \highlight{35\,\%} of them are starbursts, \highlight{40\,\%} are main-sequence galaxies and the remaining 25\,\% are located in the star-formation valley. Our sample includes one active galactic nucleus and six submillimeter sources, as revealed from ancillary X-ray and submillimeter \highlight{photometry}. The high dust extinctions combined with the flux boosting from H$\alpha$ emission explain why these sources are relatively bright Spitzer galaxies and yet unidentified in the ultradeep UltraVISTA near-infrared images.

 \end{abstract}

\section{Introduction} \label{sec:intro}
Over the past decade, deep near-/mid-infrared galaxy surveys have been used to study galaxy evolution up to high redshifts with ever-increasing statistics (e.g., \citealt{laigle2016, deshmukh2018, weaver2022}). A number of these galaxy surveys have been conducted with the Infrared Array Camera \citep[IRAC; ][]{fazio2004} on board the Spitzer Space Telescope \citep{werner2004}, in different cosmological fields \citep[e.g.,][]{ashby2015,ashby2018}. Other galaxy surveys have been carried out from the ground \highlight{at near-infrared wavelengths}, with infrared telescopes such as the VISTA telescope \citep[e.g.][]{mccracken2012,jarvis2013}.  A key factor to make these galaxy surveys successful has been the availability of ancillary photometry of matching depth at optical wavelengths, either from the Hubble Space Telescope (HST) or ground-based telescopes. This multiwavelength coverage allows for properly constraining the galaxy spectral energy distribution (SED) properties and particularly deriving secure photometric redshifts.

Not all infrared galaxies are detected at optical wavelengths, though, even when deep images are available \highlight{(for recent studies on this population, see, e.g., \citealt{barrufet2023,perez2023,smail2023})}. Some galaxies can be relatively bright in the Spitzer mid-infrared images ($\lambda > 3 \, \rm \mu m$), but faint even at near-infrared ($\lambda=1$--$3 \, \rm \mu m$) wavelengths.  A number of studies have been devoted to studying the nature of these very red sources. They concluded that these galaxies are mostly at $z=3$--6, with a subset of candidates for higher redshifts \citep[e.g.,][]{caputi2012,wang2019}. Although a minority amongst all high-redshift galaxies, mid-infrared-bright, near-infrared-faint galaxies account for half of the most massive galaxies at $z=4$-6 \citep{caputi2015}. From their follow-up at sub/millimeter wavelengths, some of these galaxies have been found to be dusty star-forming galaxies \citep{ikarashi2017,shu2022,smail2023,zavala2023}, which contribute significantly more to the cosmic star-formation rate density than equally massive UV-bright galaxies at $z>3$ \citep{zavala2021}.

More difficult to constrain is the small percentage of Spitzer sources for which no counterpart at all has been found in deep near-infrared images. The advent of the James Webb Space Telescope (JWST) allows us now to investigate the nature of these sources by providing, for the first time, ultradeep multiwavelength coverage at infrared wavelengths. Recent results have shown that in these red sources, \highlight{which have no optical counterparts}, the enhanced mid-infrared with respect to near-infrared flux is explained by a combination of factors, typically high dust extinction combined with an intermediate/high redshift \citep{barrufet2023,perez2023}.

\highlight{Previous to the launch of the JWST, it has been demonstrated how} the  presence of emission lines can also potentially produce such red colors, although only a few cases of this kind have been found to date \citep{alcalde2019}. Indeed emission lines can produce a significant enhancement even in broadband fluxes if their equivalent widths are sufficiently high. This is particularly the case for H$\alpha$~($\lambda 6563$\,\AA). A number of works have successfully exploited the Spitzer/IRAC images to identify intense H$\alpha$ emitters at $z\approx4$--5 from flux excess in the IRAC 3.6 $\mu$m band \citep{smit2016,caputi2017,faisst2019}, with rest-frame equivalent widths ($\rm EW_{0}$) ranging from $\sim100$--10,000\,\AA. These studies are biased toward galaxies well detected in the near-infrared, as good constraints on the continuum blueward of the 3.6$\mu$m band are necessary to identify any flux excess. No particular connection between line emitters and dropout galaxies has been derived from these studies.

In this work, we exploit the JWST/NIRCam imaging in the PRIMER survey \citep{primer} together with ancillary HST Advanced Camera for Surveys (ACS) and HST Wide Field Camera 3 (WFC3) imaging, to follow up IRAC-selected galaxies from the Spitzer Matching survey of the UltraVISTA ultradeep Stripes (SMUVS; \citealt{ashby2018}) that are undetected in the UltraVISTA DR5 $H$- and $K_{\mathrm{s}}$ bands. 
The improved depth and wavelength coverage of the new JWST data enables us to reveal the true nature of these so far elusive galaxies. 

This paper is organized as follows. In Sect.\,\ref{sect:data} we describe the ground- and space-based imaging used in this work, and Sect.\,\ref{sect:source_selection} describes our initial selection of dropout candidates. \highlight{In Sect.\,\ref{sec:jwst_phot}, we describe the derived HST and JWST aperture photometry for the JWST-detected counterparts of the dropout candidates.} In Sect.\,\ref{sect:properties}, we describe the subsequent SED fitting of our sample and present their general properties. In Sect.\,\ref{sect:halpha} we present our results on the H$\alpha$ line complex equivalent widths and SFRs. Finally, in Sect.\,\ref{sect:conclusion} we summarize our findings and discuss our conclusions. We adopt a cosmology with $\rm{H_0 = 70\ km\ s^{-1}\ Mpc^{-1}}$, $\Omega_{\rm m}=0.3,$ and $\Omega_\Lambda=0.7$. All magnitudes and fluxes are total, with magnitudes referring to the AB system \citep{okegun1983}. Stellar masses correspond to a \citet{chabrier2003} initial mass function (IMF). 

\section{Data} \label{sect:data}

\subsection{Ground-based and Spitzer/IRAC imaging}
Our initial search for $H-$ and $K_{\rm s}$-band dropout galaxies is based on the UltraVISTA VIRCAM imaging \citep{mccracken2012} and Spitzer \citep{werner2004} IRAC \citep{fazio2004} 3.6 and 4.5 $\mu$m imaging from the SMUVS program \citep{ashby2018} in the COSMOS field \citep{scoville2007}. 

\highlight{The UltraVISTA data considered in this work correspond to the fifth data release (DR5), which contains ultradeep imaging in three stripes on the sky taken with the VIRCAM instrument on the VISTA telescope. Over the PRIMER region considered in this work, the data reach average $5 \sigma$ depths of $\sim 26$--27 mag in the $Y$, $J$, $H$, $K_{\rm s}$ and $NB118$ bands. We performed source extraction on these data ourselves as described in Sect.\,\ref{sect:source_selection}.}

\highlight{The SMUVS program conducted observations to map three UltraVISTA ultradeep stripes, over a total area of 0.66 deg$^2$. We make use of the existing IRAC 3.6 and 4.5 $\mu$m catalog from \citet{ashby2018}. Source detection and aperture photometry measurements for this catalog were conducted using point spread function (PSF)-fitting techniques implemented with \textsc{StarFinder} \citep{diolaiti2000}. Source positions were identified on both IRAC mosaics separately, and subsequently combined into a single two-band catalog using a 1$\arcsec$ matching radius, whilst retaining single-detected sources, as well for completeness. Finally, the catalog contains a total of $\sim 350,000$ galaxies down to 4$\sigma$ limits of roughly 25.0 AB mag in both IRAC bands. Over the PRIMER area, the IRAC data reach average $5 \sigma$ depths of $\sim$ 26 mag.}

\subsection{HST and JWST imaging}
In this work, we made use of archival HST and newly obtained JWST imaging to uncover the true nature of potential $HK_{\rm s}$-band dropout galaxies. We have used the JWST/NIRCam data collected during the recent GO Cycle 1 program PRIMER (PID: 1837; PI: James Dunlop; 
\citealt{primer}). PRIMER observations are conducted in eight NIRCam broad and medium bands - F090W, F115W, F150W, F200W, F277W, F356W, F410M, and F444W, and two MIRI bands, i.e. F770W and F1800W, over two CANDELS legacy fields \citep{grogin2011,koekemoer2011} in COSMOS and the Ultra-Deep Survey \citep{cirasuolo2007, lawrence2007}. Here, we consider all observations of target COSMOS-2 \highlight{, the first-released subsection of the full PRIMER target area,} taken up until 2023 January 6, with a total integration time of 36.1 hr covering an area of $\sim$94 arcmin$^2$. We did not utilize the available MIRI data as they are offset from the NIRCam data, nor did we use NIRCam/F150W-band imaging as it had not been taken yet. 

We have processed the JWST/NIRCam data in the following way. First, we retrieved the level-2 data products from the MAST archive \footnote{Data available at \dataset[10.17909/bysp-ds64]{https://archive.stsci.edu/doi/resolve/resolve.html?doi=10.17909/bysp-ds64}} and processed them with the \textsc{grizli} pipeline (version 1.7.7; \citealt{grizli}). We took particular care when introducing additional corrections for the NIRCam photometric zero-points relative to the Calibration Reference Data System pipeline mapping (pmap) \textsc{1041}, including variations across the detector itself \citep{nircamcallibration}. These were found to be consistent with the efforts of other groups \citep{boyer2022, nardiello2022}. We additionally introduce corrections and masking to account for the detrimental effect of the cosmic rays and stray light in the detector \citep[see][]{bradley2022}. More detailed descriptions of this procedure are presented in \citet{kokorev2023} and \citet{valentino2023}, and will be further expanded upon in Brammer et al. (2024, in preparation). 

We supplement the JWST observations with the available HST optical and near-infrared data available in the Complete Hubble Archive for Galaxy Evolution \citep{kokorev2022}. These data include imaging in the HST/WFC3 F125W, F140W, and F160W and HST/ACS F435W, F606W, and F814W bands. We do not include the available data in the WFC3/F105W and ACS/F475W bands as these cover only a small fraction of the PRIMER region.
All available JWST and HST exposures were aligned to the Gaia DR3 \citep{gaia2023} astrometry, co-added, and drizzled \citep{fruchter2002} onto two sets of grids. We used a $0\farcs02$ pixel scale for the short-wavelength NIRCam bands and $0\farcs04$ for all the remaining JWST and 
HST filters. For comparison, the FWHM of the NIRCAM PSF adopted in this work is $0\farcs03$ in the F090W band.

\begin{figure}
    \plotone{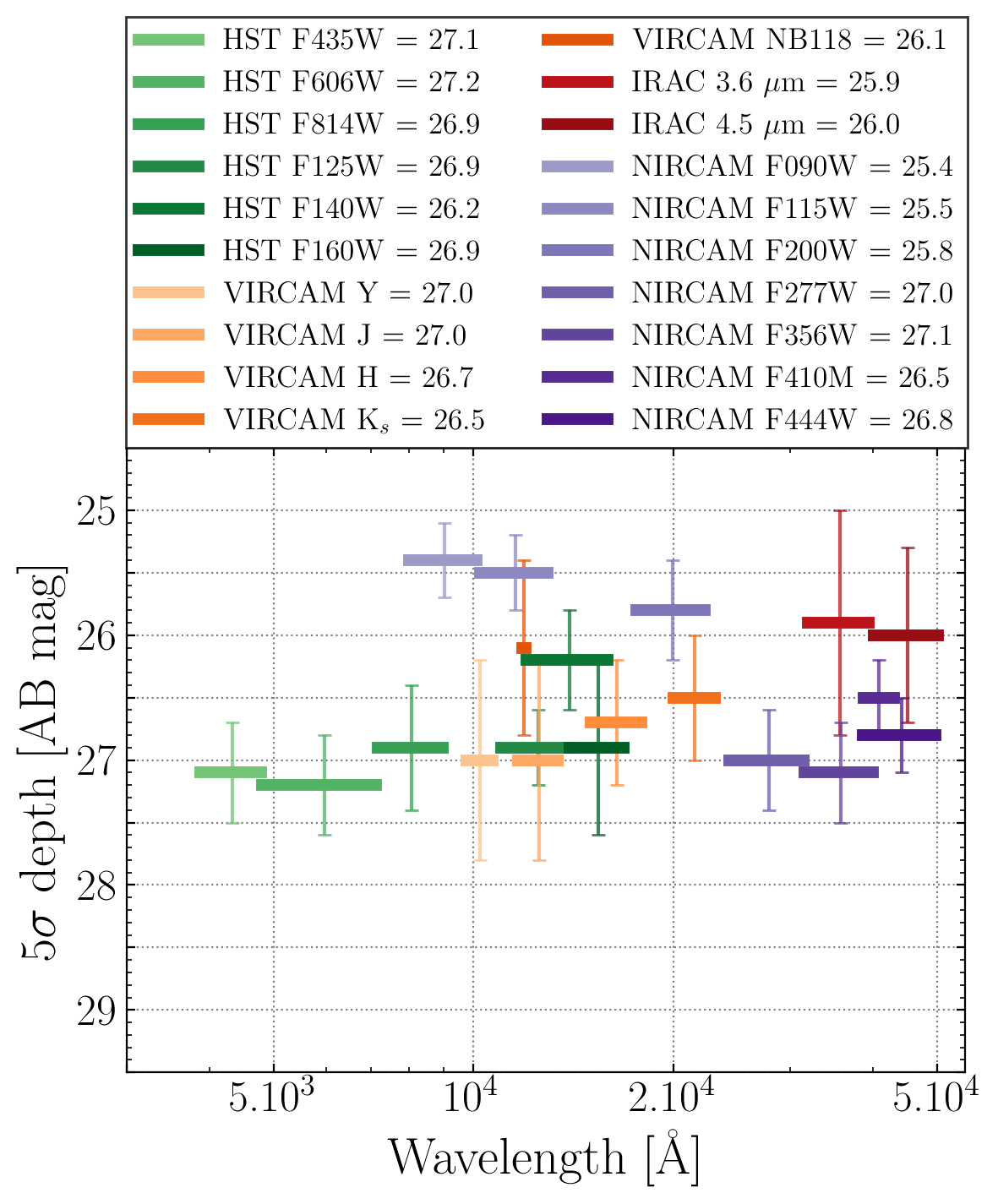}
    \caption{Depth (5$\sigma$) and associated uncertainty in each filter considered throughout this work, measured as the median and standard deviation of the local background in 30 by $30\arcsec$ regions over the PRIMER field, respectively. In each region, the local background is the standard deviation of the flux measurement in empty $2\arcsec$ circular diameter apertures. The horizontal width of each datapoint indicates the FWHM for the filter. \label{fig:depth}}
\end{figure}

The depth in each band considered throughout this work is shown in Fig.\,\ref{fig:depth}, measured from $2\arcsec$ empty apertures in the field. 

\section{Initial sample selection from IRAC and VIRCAM imaging} \label{sect:source_selection}
Our analysis is based on a selection of IRAC-detected sources in the PRIMER region that are undetected in the UltraVISTA DR5 VIRCAM $H$- and/or $K_{\rm s}$-band images. 

\highlight{We created from scratch an initial UltraVISTA DR5 catalog to crossmatch with the IRAC combined 3.6 and 4.5 $\mu$m catalog from \citet{ashby2018}. We performed source detection on the UltraVISTA $H$- and $K_{\rm s}$-band DR5 images separately using \textsc{SourceExtractor} \citep{bertin1996}. We adopted a $1.5\sigma$ detection threshold over five contiguous pixels. We used \textsc{SourceExtractor} in dual-image mode to measure the supporting $Y$, $J$, and NB118 fluxes at the locations of the $H$- and $K_{\rm s}$-detected sources. To ensure we catch all the light for the flux measurements, we extracted both the \textsc{SourceExtractor} 'mag\_auto' and the aperture magnitude measured from $2\arcsec$ apertures. Subsequently, for each band individually, we chose between these flux measurements based on the local $10 \sigma$ depth, such that we favor aperture fluxes for faint sources. The UltraVISTA images are not PSF-matched, but this poses no concern as we derive aperture flux corrections for each filter separately from the curve of growth of bright stars in the field. We applied corrections for galactic dust extinction using the \citet{schlafly2011} dust maps with the \citet{fitzpatrick1999} reddening law.}

\highlight{In total, the PRIMER region considered in this work contains 11,651 and 10,633 $H$- and $K_{\rm s}$-detected sources, respectively, and 7057 IRAC-detected galaxies from the \citet{ashby2018} catalog. We crossmatched all UltraVISTA sources that have $>2\sigma$ flux detections in the $Y$, $J$, $H$, $K_{\rm s}$, and NB118 bands with the IRAC galaxies within a $1\arcsec$ search radius. We kept all IRAC sources that are unmatched with any UltraVISTA source within this search radius, yielding 1441 potential $K_{\rm s}$-band dropout candidates and 148 potential $H$-band dropout candidates.} 

\highlight{We proceeded to carefully inspect these candidates through various measures. First, we used the JWST/NIRCam F444W imaging to check the validity of the IRAC sources, as any IRAC source is expected to be visible in these data of superior depth and resolution. We identified the JWST sources by running \textsc{SourceExtractor} on the F444W image with a detection threshold of 1.5$\sigma$ over three contiguous pixels and a} \verb|gauss_3.0_3x3.conv| convolution mask. \highlight{The IRAC sources were crossmatched with the F444W-detected sources in a $1\arcsec$ search radius, and any IRAC source with no JWST source located in its aperture was removed from the sample, that being 32 and 3\,\% of the initial $K_{\rm s}$- and $H$-band dropout candidate samples, respectively. From visual inspection, we determined that 93\,\% of these removed candidates are falsely detected residuals of bright IRAC galaxies.} 

\highlight{Our second check was to crossmatch the dropout candidates with the HSC bright star masks \citep{aihara2022} and the Gaia DR3 star catalog \citep{gaia2023}. Any IRAC source located within the HSC masks or within a $3\arcsec$ radius of a Gaia source was removed from the sample, that being 16 and 41\,\% of the initial $K_{\rm s}$- and $H$-band dropout candidate samples, respectively. This was done to avoid flux contamination from the bright, nearby star due to the broad IRAC PSF.} 

\highlight{For our third and final check, we removed any dropout candidate that has significant UltraVISTA fluxes. For example, an intrinsically faint galaxy with strong emission lines may be undetected in the $K_{\rm s}$ image but be perfectly visible in the $Y$ or $J$ band. To measure the VIRCAM fluxes of the dropout candidates, we performed forced aperture photometry on the VIRCAM images at the IRAC positions, using the \textsc{Python} modules \textsc{Astropy} (version 5.0.4; \citealt{astropy}) and \textsc{Photutils} (version 1.4.1; \citealt{photutils}), as contrary to \textsc{SourceExtractor}, one can directly supply a list of source locations to this software. The photometry was measured on the $Y$, $J$, $H$, $K_{\rm s}$, and NB118 bands in $2\arcsec$ apertures. We derived the flux errors as the standard deviation of the $3\sigma$-clipped (five iterations) background flux distribution, measured from empty apertures placed over a 10 by 10$\arcsec$ region around the source. We required a minimum error of 0.1 magnitude error, corrected the fluxes to total, and applied galactic extinction corrections.}

\highlight{In principle, we discarded all $K_{\rm s}$-band dropout candidates that have $> 2\sigma$ flux measurements in any of the $Y$, $J$, $H$, $K_{\rm s}$, and NB118 bands. Similarly, we considered the $Y$, $J$, $H$, and NB118 flux measurements for the $H$-band dropout candidates. However, we carefully visually inspected every dropout candidates discarded by this measure, as the flux errors can be affected by a crowded background, and decided to keep four $K_{\rm s}$-  and four $H$-band dropout candidates in our final dropout candidate sample for this reason. Moreover, upon visual inspection of the dropout candidates with no significant VIRCAM fluxes, we find that between both the $K_{\rm s}$- and $H-$ dropout candidates, roughly a third are IRAC residual detections. These detections survived our earlier checks because they coincidentally match with a very faint F444W-detected source. However, the focus of our work are isolated IRAC galaxies with a direct correspondence to a JWST-detected source, so we discard these residual detections from the final sample. In addition, we discarded another $\sim 20$\,\% of the galaxies with no significant VIRCAM flux measurements because we visually confirmed the presence of a convincing source in these bands, and another $10$\,\% because the VIRCAM and IRAC imaging are highly contaminated by the presence of a nearby bright source.} 

\highlight{After implementing the abovementioned checks, our final sample consists of 18 and 8 $K_{\rm s}$- and $H$-band dropout candidates, respectively. This amounts to 0.4\,\% of all IRAC-detected sources in PRIMER from the \citet{ashby2018} catalog. The IRAC magnitudes of the final dropouts range between 23.1--27.0 mag, with the signal-to-noise ratio (S/R) varying between 3.5 and 10.9. For comparison, considering all IRAC-detected sources over the PRIMER field, the median 3.6 and 4.5 $\mu$m magnitudes are 23.6 and 23.7 with typical S/Ns of $\sim 40$.}

\section{HST and JWST photometry of the dropout candidates} \label{sec:jwst_phot}
\highlight{Now that we have selected a sample of apparent dropout galaxies from VIRCAM and IRAC data, we turn to the JWST and HST imaging to study these sources in detail.} 

Given the superior resolution of the JWST imaging, instead of using the IRAC priors to measure the NIRCam aperture photometry, we detected sources directly on the F277W/F356W/F444W stacked image using \textsc{SourceExtractor}. Similar to the extraction of sources on the singular F444W image in the previous section, we required a detection threshold of 1.5$\sigma$ over three contiguous pixels and convolved the data with a \verb|gauss_3.0_3x3.conv| mask. In addition, we performed identical runs on the F200W and F277W images alone, as some of the dropout candidates suffer from source blending in the longest wavelength JWST channels. \highlight{We crossmatched our JWST-detected sources with the IRAC positions of the dropout candidates and retained all within a $1\arcsec$ radius of the candidates, i.e., all that would fall within the IRAC $2\arcsec$ apertures adopted throughout this work. In 11 out of the 26 UltraVISTA dropout candidates, the JWST imaging shows multiple counterparts. For each counterpart, we measure the HST and \textit{JWST } aperture fluxes individually. In total, the crossmatching with the IRAC priors yields 51 JWST-detected galaxies.}

To measure the HST and JWST photometry of the crossmatched galaxies, we again performed forced aperture photometry with \textsc{Astropy} and \textsc{Photutils}. \highlight{We emphasize that we did not create a fully realized JWST-detected source catalog over the PRIMER field for this study}. We measured fluxes in both $0\farcs5$ and $1\arcsec$ diameter circular apertures at the positions of the JWST-detected sources on the following 13 optical and near-infrared bands, that together span the wavelength range 0.4--4.4 $\mu$m: \highlight{JWST/NIRCAM F090W, F115W, F200W, F277W, F356W, F410M, and F444W, and HST/WFC3 F125W, F140W,and  F160W, and HST/ACS F435W, F606W, and F814W.} We decided between the $0\farcs5$ and $1\arcsec$ diameter apertures on a source-by-source basis from careful visual inspection of the NIRCam images. Similar to the routine described in the previous section, we derived the flux errors as the standard deviation of the $3\sigma$-clipped (five iterations) background flux distribution, measured from empty apertures placed over a 10 by 10$\arcsec$ region around the source. We required a minimum error of 0.05 magnitude error for both the HST and JWST photometry. Assuming a point-source morphology, the HST aperture fluxes were corrected to total using the median curve of growth of stars in the field.  Aperture corrections for the JWST fluxes were estimated using the \textsc{WebbPSF} \footnote{WebbPSF is available at \url{https://webbpsf.readthedocs.io/en/latest/}} software instead, as virtually all stars in PRIMER-COSMOS are saturated in the NIRCam imaging. We applied corrections for galactic dust extinction. In bands with nondetections, we adopted $3\sigma$ flux upper limits derived from the empty aperture fluxes.

\section{Results}
\subsection{SED fitting based on HST and JWST photometry} \label{sect:properties}  

\highlight{In this section, we describe how we derived the redshifts and physical parameters based on the HST and JWST photometry of the 26 dropout candidates. To ensure the validity of the photometric redshifts for these sources, we ran three separate redshift codes: \textsc{LePhare}, \textsc{EAZY}, and \textsc{Prospector}. The latter two codes are strictly used to verify the redshift results produced by \textsc{LePhare}. We describe the routines adopted for the three codes in the following paragraphs.} 

First, the $\chi^2$-fitting algorithm \textsc{LePhare} \citep{arnouts1999,ilbert2006} uses a set of galaxy templates created from stellar population synthesis to find the best-fit galaxy solution to the observed photometry using a $\chi^2$-minimization technique, thereby deriving the physical parameters of the observed source. The galaxy models adopted in this work were sampled from the GALAXEV library \citep{bruzual2003}, \highlight{which consists of evolutionary stellar population synthesis models constructed with isochrones and a collection of theoretical and (semi-)empirical stellar spectra.}

To compile our galaxy models, we adopt different star formation histories (SFHs): a single stellar population and two parametric SFHs, i.e. the star formation rate (SFR) is an analytical function of time. For the latter, we adopted an exponentially declining SFH ($\rm{SFR} \propto e^{-t/\tau}$) and a delayed exponentially declining ($\rm{SFR} \propto te^{-t/\tau}$), using the range of star formation timescales $\tau = 0.01, 0.1, 0.3, 1.0, 3.0, 5.0, 10.0$, and 15 Gyr. We considered two metallicities: solar ($Z=Z_\odot$) and subsolar ($Z=0.2Z_\odot$). This results in 36 different combinations of the SFH and metallicity for our considered SED models. 

For dust attenuation, we adopted the \citet{calzetti2000} reddening law and left the color excess as a free parameter between $E(B-V)=0$ and 1, with a step of 0.1. For the galaxy ages, we adopted a grid of values spanning 0.01--13.5 Gyr. \highlight{We applied emission lines to the galaxy templates using \textsc{LePhare}, which are modeled using scaling relations from \citet{kennicutt1998} between the UV luminosity and [OII] line. In this treatment, the following emission lines are considered with individual ratios compared to the O[II] line: Ly$\alpha$, H$\alpha$, H$\beta$, [OIII], and O[II]~($\lambda 4959,\lambda5007$\,\AA). The intensity of the lines is scaled with the intrinsic UV luminosity of the galaxy. We refer the reader to \citet{ilbert2009} for a detailed description of the emission-line implementation in \textsc{LePhare}.} Absorption of emission at wavelengths shorter than rest-frame $912$\,\AA\ by the intergalactic medium (IGM) was implemented following \citet{madau1995}. Finally, any templates that produce fluxes exceeding the 3$\sigma$ flux upper limits in nondetected bands are rejected by \textsc{LePhare}.

If our candidates drop out of the UltraVISTA bands because these sample the Lyman break, the sources could be at redshifts as high as $z\sim17$. Therefore, we performed two runs with different redshift ranges: one spanning $z=0$--9 and one extending to $z=17$. Similarly, the dropout nature could be caused by extreme dust reddening, such that we ran one additional SED fit with $z=0$--9 but $E(B-V)$ extending to 1.5. We subsequently compared the reduced $\chi^2$ of the three fits for each source, where we adopt the results from the run that yields the smallest $\chi^{2}_{\nu}$ value. \highlight{We adopted this approach in an effort to mitigate the effects of photometric redshift degeneracy, which red galaxies are particularly prone to.}

\highlight{Second, we used the \textsc{Python} version of \textsc{EAZY} \citep{brammer08}. \textsc{EAZY} employs a set of nonnegative linear combinations of basic templates constructed with the Flexible Stellar Population Synthesis (FSPS; \citealt{conroy2009, conroy2010}) models. In particular, this analysis used the \textsc{corr\_sfhz\_13} subset of models. These contain redshift-dependent SFHs, which at a given redshift are restricted by the age of the Universe. In addition, the maximum allowed dust attenuation is tied to a given epoch. We also included the best-fit template to the JWST-observed extreme emission-line galaxy at $z=8.5$ (ID4590) from \citet{carnall22}, which was renormalized in alignment with the FSPS models. This was done in order to model potentially large equivalent-width emission lines.} 

\highlight{We adopted the \textsc{EAZY} template error function to accommodate additional uncertainty related to unusual stellar populations, using the default value of 0.2 for the template error. We considered a redshift range of $z=0.01$--17 for this run. Because \textsc{EAZY} does not allow for the same treatment of nondetected bands as \textsc{LePhare}, we instead adopted the $1\sigma$ flux upper limit as a flux measurement for the nondetected bands in the fitting process.} 

\highlight{Lastly, we made use of the Bayesian inference code \texttt{Prospector} \citep{johnson2021}. Similarly to \textsc{EAZY}, its stellar population models are based on the FSPS code, but \textsc{Prospector} rather builds the galaxy SED models on the fly using the \textsc{Python}-FSPS bindings. \textsc{Prospector} is well-known for its ability to model complex, nonparametric SFHs (see, e.g., \citealt{leja2017,tachella2022}). However, given that our intention is to generate a photometric redshift with \textsc{Prospector} in order to verify our \textsc{LePhare} results, we stick to a simple delayed exponentially declining parametric SFH.} 

\highlight{The parametric model involves seven free parameters, for which we chose broad priors comparable to the \textsc{LePhare} run. For each parameter, we used the default prior shapes with the following ranges: the redshift $z=0$--17, the formed stellar mass $M_* = 10^6$--$10^{12} M_\odot$, the metallicity $\log(Z/Z_\odot) = -2$--0.19, the e-folding time $\tau_{\rm SF} = 0.001$--15 Gyr, and the age $t_{age} = 0.001$--13.8 Gyr. We modeled diffuse dust attenuation following \citet{calzetti2000} with $\tau_{\rm dust} = 0$--4. Lastly, we implemented IGM absorption following \citet{madau1995} and nebular emission, using the default parameters. Finally, \textsc{Prospector} also treats nondetections differently, such that we set the flux to zero and the flux uncertainty as the $1\sigma$ upper limit in the corresponding bands.}

\subsubsection{One-to-one comparison between IRAC- and JWST-detected galaxies}
\highlight{In Sect.\,\ref{sec:redshifts}, we discuss the results from our SED fitting on the HST and JWST photometry of all JWST-detected galaxies that are counterparts to our 26 UltraVISTA dropout candidates. As we remind the reader, the dropout candidates crossmatch with 51 JWST-detected galaxies. Therefore, for each system, the one-to-one correspondence between the IRAC- and JWST-detected galaxies varies, as we discover from the combined inspection of the JWST source detection, the \textsc{LePhare}-derived photometric redshifts, and the image cutouts of the 26 UltraVISTA dropout candidates in the HST, JWST, IRAC, and ancillary ground-based bands (VIRCAM and HSC). In the interest of clarity and structure, we therefore first discuss the subclassification of the 26 dropout candidates before we present the photometric redshift evaluation in the next section. }

The dropout image cutouts of the candidates can be found in the Appendix in Figures\,\ref{fig:stamps_secure_1}, \ref{fig:stamps_secure_2}, \ref{fig:stamps_deboosted_1}, \ref{fig:stamps_deboosted_2}, and \ref{fig:stamps_associated}.

\highlight{First, we have checked if the UltraVISTA dropout candidates are true dropout galaxies. Following the strategy applied to the VIRCAM photometry, we checked if the JWST-detected counterparts have $>2\sigma$ flux measurements in the HST and JWST bands blueward of the NIRCAM/F200W and WFC3/F160W bands for the $K_{\rm s}$- and $H$-band dropout candidates respectively. We find that out of the 26 dropout candidates, 4 systems display no significant flux in these bands; 2 $K_{\rm s}$- and 2 $H$-band dropout candidates are true dropout galaxies. }

Second, in \highlight{15 out of the 26 candidates}, the superior resolution of the HST and JWST imaging compared to the IRAC data enables us to identify multiple sources residing in the $2\arcsec$ diameter IRAC apertures. In fact, only \highlight{11 out of 26 candidates} appear as one galaxy in JWST, i.e., there is a one-to-one correspondence with the IRAC detection. The other \highlight{15 candidates} are composed of multiple sources. Among these systems, \highlight{10} cases consist of a primary galaxy with one or more nonassociated foreground sources contaminating the aperture photometry, boosting the flux of the primary source. \highlight{We identify the primary galaxy as the brightest source in the system considering the NIRCAM/F444W flux, which coincidentally is also closest in angular separation to the IRAC source position. The contaminating sources in each system emit less than 10\% of the F444W flux of the primary galaxy, with the exception of two systems. However, in these two systems, the F444W fluxes and photometric redshifts of the contaminates are significantly lower than those of the main galaxy considering the error bars, such that we can safely deem them to be not associated. By contrast}, the remaining \highlight{five} candidates are systems of galaxies that are partially associated with each other. \highlight{This means that multiple individual components in these systems have closely related redshifts and/or comparable F444W fluxes (i.e. $>10\%$ of the F444W flux of the brightest galaxy in the system), so that their appearance in the IRAC aperture is either not simply the result of superposition, or their contribution to the IRAC flux cannot be dismissed.} Throughout this paper, we refer to these categories as \highlight{(11) secure galaxies, (10) deboosted galaxies, and (5) associated galaxies.}  

\highlight{As mentioned before, the crossmatching with the IRAC priors yields 51 JWST-detected galaxies, although the contaminating sources of the deboosted sample are omitted from our analysis, as these low-redshift foreground galaxies are beyond the focus of our work. Therefore, the total number of JWST-detected galaxies we discuss throughout this work is 35.}

\subsubsection{Evaluation of the HST- and JWST-derived photometric redshifts}\label{sec:redshifts}
\highlight{Now that we have explained how the UltraVISTA dropout candidates are classified into different categories based on their JWST-detected counterparts, we discuss the redshift results from the three separate SED fitting routines performed with \textsc{LePhare}, \textsc{EAZY}, and \textsc{Prospector}.} 

\highlight{Table \ref{tab:redshifts} shows the best-fit photometric redshifts obtained for each of the relevant 35 JWST-detected galaxies, as well as the source position, [F160W-F444W] color and size of the aperture used to measure the HST and JWST photometry.} We distinguish between candidates originally identified as $H$- and $K_{\rm s}$-band dropouts candidates by means of their ID. \highlight{We note that due to the close proximity and therefore overlap in sources of systems PD-H-a-1 and PD-H-a-2, we consider them together as one system PD-H-a-1/2 throughout this work}. We also group the sources in Table\,\ref{tab:redshifts} by their respective classes (secure, deboosted, and associated) for clarity. \highlight{The \textsc{LePhare} and \textsc{EAZY} photometric redshifts correspond to the model that minimizes $\chi^2_{\nu}$; for \textsc{Prospector}, the median of the probability distribution function of the redshift, PDF($z$), is reported. In all three cases, the uncertainty on the photometric redshift corresponds to the \nth{16} and \nth{84} percentiles. }

\begin{figure*}
    \plotone{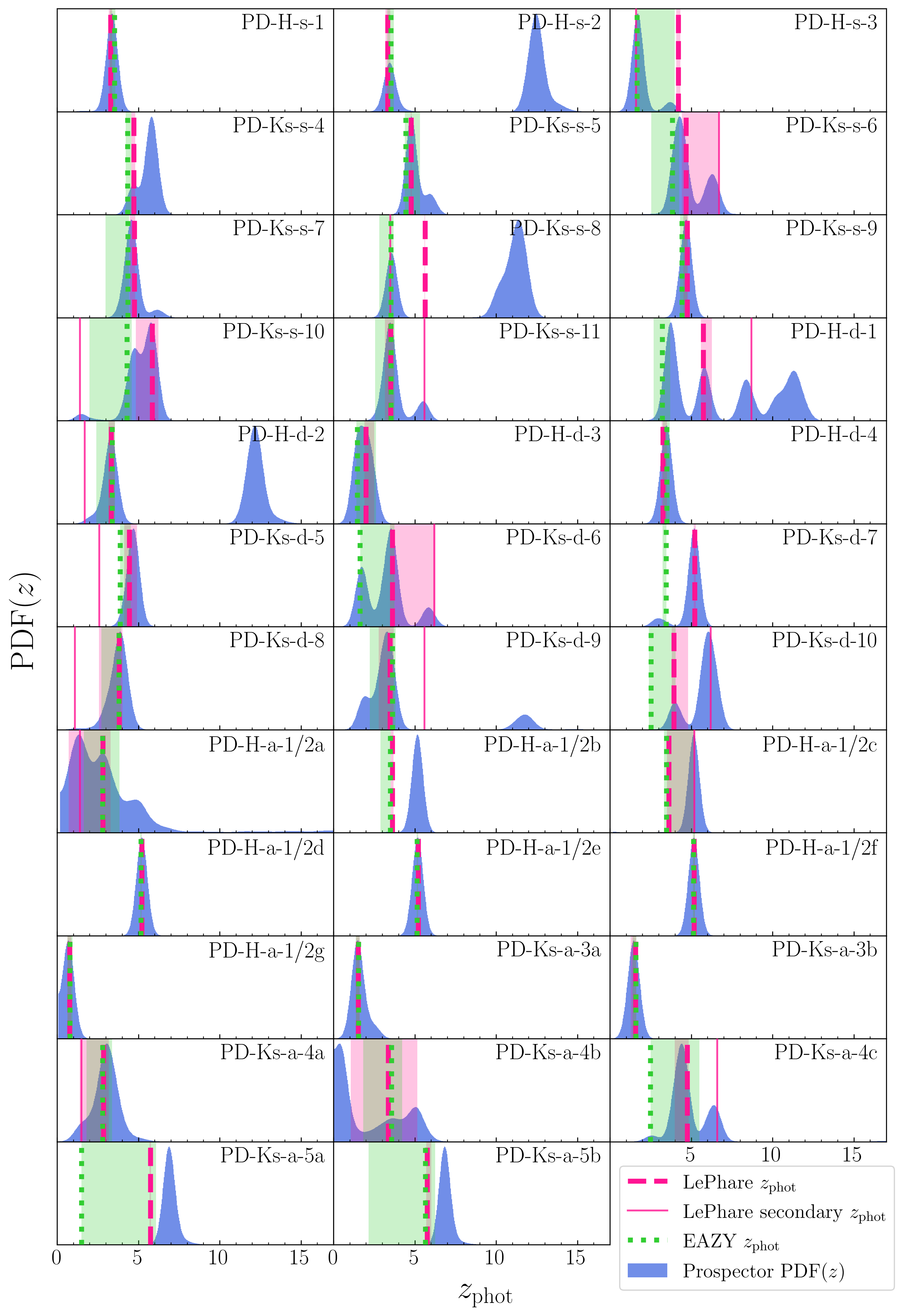}
    \caption{Photometric redshift distribution of the dropout candidates derived from SED fitting the HST and JWST photometry. Each panel shows one of the 35 JWST-detected counterparts to the UltraVISTA dropout candidates considered in this work, depicting the probability density function of the redshift produced by \textsc{Prospector} in blue. The best-fit photometric redshift solutions from \textsc{LePhare} and \textsc{EAZY} are shown with pink dashed and green dotted vertical lines, respectively. We also show for the \textsc{LePhare} and \textsc{EAZY} redshift solutions the \nth{16} and \nth{84} percentiles by means of correspondingly colored shaded regions. Finally, the secondary redshift from \textsc{LePhare}, when available, is shown in a solid pink vertical line. \label{fig:redshift_individual}}
\end{figure*}

\highlight{We visualize our redshift results in Fig.\,\ref{fig:redshift_individual}, where we show for each JWST-detected galaxy the PDF($z$) from \textsc{Prospector}, as well as the best-fit redshifts and uncertainty ranges from \textsc{LePhare} and \textsc{EAZY}. We also indicate the secondary redshift form \textsc{LePhare}, i.e., the competing model that minimizes the $\chi^2_{\nu}$ second-best.}

\begin{deluxetable*}{lcccccccc}
 \centerwidetable 
 \tabletypesize{\footnotesize}
\tablecaption{Photometric redshifts of JWST-detected galaxies located within the IRAC apertures \label{tab:redshifts}}
\tablewidth{0pt}
\tablehead{
\multicolumn{2}{c}{ID} & \colhead{R.A.} & \colhead{Decl.} & \colhead{[F160W-F444W]} & \colhead{Aperture} & \multicolumn{3}{c}{Photometric Redshift} \\
\multicolumn{2}{c}{} & \colhead{} & \colhead{} & \colhead{(AB mag)} & \colhead{$\arcsec$} & \colhead{LePhare} & \colhead{EAZY} & \colhead{Prospector}}
\startdata
PD-H-s-1 & & 10:00:30.42 & +02:26:17.16 & $3.6 \pm 0.2$ & 0.5 & $3.28^{+0.10}_{-0.10}$ & $3.53^{+0.06}_{-0.28}$ & $3.39^{+0.09}_{-0.12}$\\
PD-H-s-2 & & 10:00:27.98 & +02:25:29.62 & $3.8 \pm 0.6$ & 1.0 & $3.32^{+0.17}_{-0.13}$ & $3.53^{+0.18}_{-0.22}$ & $12.34^{+0.38}_{-8.84}$\\
PD-H-s-3 & & 10:00:47.04 & +02:19:56.52 & $1.5 \pm 0.2$ & 0.5 & $4.21^{+0.12}_{-0.13}$ & $1.67^{+2.33}_{-0.01}$ & $1.73^{+0.11}_{-0.09}$\\
PD-Ks-s-4 & & 10:00:36.19 & +02:15:50.76 & $2.4 \pm 0.5$ & 0.5 & $4.71^{+0.09}_{-0.15}$ & $4.34^{+0.22}_{-0.13}$ & $5.74^{+0.26}_{-0.1.00}$\\
PD-Ks-s-5 & & 10:00:33.86 & +02:20:18.15 & $2.4 \pm 0.2$ & 0.5 & $4.78^{+0.04}_{-0.02}$ & $4.45^{+0.87}_{-0.07}$ & $4.80^{+0.95}_{-0.06}$\\
PD-Ks-s-6 \tablenotemark{b} & & 10:00:24.15 & +02:20:05.40 & $4.1 \pm 1.8$ & 1.0 & $4.68^{+2.03}_{-0.25}$ & $3.83^{+0.42}_{-1.29}$ & $4.48^{+1.78}_{-0.46}$\\
PD-Ks-s-7 & & 10:00:40.80 & +02:26:37.37 & $3.5 \pm 0.8$ & 0.5 & $4.75^{+0.06}_{-0.28}$ & $4.35^{+0.07}_{-1.37}$ & $4.62^{+0.15}_{-0.26}$\\
PD-Ks-s-8 & & 10:00:35.34 & +02:28:26.63 & 2.9 \tablenotemark{a} & 1.0 & $5.63^{+0.00}_{-0.04}$ & $3.53^{+0.16}_{-0.73}$ & $10.97^{+0.66}_{-7.37}$ \\
PD-Ks-s-9 & & 10:00:28.73 & +02:16:01.25 & $4.9 \pm 2.6$ & 0.5 & $4.75^{+0.06}_{-0.24}$ & $4.42^{+0.33}_{-0.11}$ & $4.71^{+0.08}_{-0.17}$\\
PD-Ks-s-10 & & 10:00:48.53 & +02:27:48.77 & $2.2 \pm 0.3$ & 0.5 & $5.85^{+0.39}_{-1.00}$ & $4.30^{+0.28}_{-2.32}$ & $5.47^{+0.47}_{-0.90}$ \\
PD-Ks-s-11 \tablenotemark{b}& & 10:00:27.03 & +02:24:24.03 & $3.8 \pm 0.5$ & 1.0 & $3.51^{+0.19}_{-0.35}$ & $3.53^{+0.23}_{-0.97}$ & $3.54^{+0.90}_{-0.27}$ \\[0.2cm]
\hline 
PD-H-d-1 \rule{0pt}{1.5\normalbaselineskip} & & 10:00:41.83 & +02:25:47.02 & 3.2 \tablenotemark{a} & 1.0 & $5.73^{+0.54}_{-0.14}$ & $3.23^{+0.48}_{-0.54}$ & $6.00^{+5.16}_{-2.28}$\\
PD-H-d-2\tablenotemark{b}  & & 10:00:22.44 & +02:23:40.99 & $2.9 \pm 0.5$ & 0.5 & $3.32^{+0.23}_{-0.16}$ & $3.39^{+0.18}_{-0.97}$ & $11.74^{+0.66}_{-8.52}$\\
PD-H-d-3 & & 10:00:38.31 & +02:25:44.22 & $1.8 \pm 0.2$ & 1.0 & $1.99^{+0.51}_{-0.09}$ & $1.45^{+1.16}_{-0.04}$ & $1.91^{+0.45}_{-0.50}$\\
PD-H-d-4 \tablenotemark{b} & & 10:00:28.94 & +02:25:05.13 & $6.2 \pm 3.8$ & 1.0 & $3.25^{+0.26}_{-0.07}$ & $3.50^{+0.18}_{-0.24}$ & $3.49^{+0.12}_{-0.15}$\\
PD-Ks-d-5 & & 10:00:37.48 & +02:15:57.82 & $2.7 \pm 0.7$ & 1.0 & $4.44^{+0.47}_{-0.30}$ & $3.88^{+0.66}_{-0.00}$ & $4.75^{+0.08}_{-0.30}$\\
PD-Ks-d-6 & & 10:00:45.86 & +02:22:02.26 & $2.0 \pm 0.3$ & 0.5 & $3.62^{+2.64}_{-0.16}$ & $1.63^{+2.11}_{-0.02}$ & $3.43^{+0.32}_{-1.67}$\\
PD-Ks-d-7 & & 10:00:31.20 & +02:25:04.07 & $1.6 \pm 0.2$ & 1.0 & $5.21^{+0.06}_{-0.06}$ & $3.46^{+0.01}_{-0.22}$ & $5.19^{+0.07}_{-0.12}$\\
PD-Ks-d-8 & & 10:00:26.31 & +02:19:38.36 & $0.6 \pm 0.4$ & 0.5 & $3.82^{+0.20}_{-1.23}$ & $3.80^{+0.12}_{-1.08}$ & $3.83^{+0.34}_{-0.58}$\\
PD-Ks-d-9 & & 10:00:45.96 & +02:18:52.85 & $2.7 \pm 0.4$  & 0.5 & $3.47^{+0.24}_{-0.69}$ & $3.62^{+0.21}_{-1.38}$ & $3.22^{+0.51}_{-1.02}$\\
PD-Ks-d-10 & & 10:00:27.78 & +02:25:52.04 & 2.9 \tablenotemark{a} & 1.0 & $3.93^{+0.87}_{-0.13}$ & $2.51^{+1.52}_{-0.12}$ & $5.97^{+0.47}_{-1.72}$\\[0.2cm]
\hline 
\multirow{ 7}{*}{PD-H-a-1/2} \rule{0pt}{1.5\normalbaselineskip} & a & 10:00:37.73 & +02:16:55.07 & $0.0 \pm 1.2$ & 0.5 & $2.82^{+0.48}_{-2.11}$ & $2.79^{+1.04}_{-1.15}$ & $2.47^{+2.14}_{-1.36}$ \\
& b & 10:00:37.78 & +02:16:54.53 & $2.2 \pm 1.1$ & 0.5 & $3.61^{+0.09}_{-0.09}$ & $3.48^{+0.05}_{-0.60}$ & $5.17^{+0.03}_{-0.04}$ \\
& c & 10:00:37.75 & +02:16:54.42 & $0.7 \pm 0.5$ & 0.5 & $3.61^{+1.59}_{-0.11}$ & $3.47^{+1.68}_{-0.14}$ & $5.16^{+0.03}_{-0.04}$ \\
& d & 10:00:37.72 & +02:16:54.34 & $1.3 \pm 0.5$ & 0.5 & $5.21^{+0.00}_{-0.02}$ & $5.16^{+0.04}_{-0.03}$ & $5.20^{+0.06}_{-0.04}$\\
& e & 10:00:37.71 & +02:16:54.01 & $2.3 \pm 0.6$ & 0.5 & $5.21^{+0.00}_{-0.01}$ & $5.14^{+0.05}_{-0.02}$ & $5.18^{+0.03}_{-0.03}$ \\
& f & 10:00:37.69 & +02:16:53.56 & $2.3 \pm 0.2$ & 0.5 & $5.18^{+0.04}_{-0.01}$ & $5.14^{+0.08}_{-0.04}$ & $5.16^{+0.05}_{-0.05}$\\
& g & 10:00:37.68 & +02:16:52.74 & $-0.3 \pm 0.4$ & 0.5 & $0.77^{+0.11}_{-0.13}$ & $0.79^{+0.16}_{-0.14}$ & $0.70^{+0.13}_{-0.13}$\\
\cline{1-2}
\multirow{ 2}{*}{PD-Ks-a-3} & a & 10:00:40.04 & +02:23:16.75 & $-0.1 \pm 0.3$ & 0.5 & $1.52^{+0.10}_{-0.13}$ & $1.53^{+0.08}_{-0.17}$ & $1.52^{+0.19}_{-0.18}$\\
& b & 10:00:40.02 & +02:23:16.57 & $0.0 \pm 0.3$ & 0.5 & $1.57^{+0.05}_{-0.29}$ & $1.59^{+0.04}_{-0.20}$ & $1.47^{+0.13}_{-0.14}$\\
\cline{1-2}
\multirow{ 3}{*}{PD-Ks-a-4} & a & 10:00:45.05 & +02:24:41.74 & $0.6 \pm 0.4$ & 0.5 &$2.86^{+0.33}_{-1.46}$ & $2.80^{+0.59}_{-0.98}$ & $3.02^{+0.61}_{-0.76}$\\
& b & 10:00:45.10 & +02:24:41.38 & $0.2 \pm 0.7$ & 0.5 & $3.36^{+1.81}_{-2.30}$ & $3.56^{+0.66}_{-1.73}$ & $1.38^{+3.55}_{-0.99}$\\
& c & 10:00:45.07 & +02:24:40.66 & 1.8 \tablenotemark{a} & 0.5 & $4.76^{+0.09}_{-0.77}$ & $2.49^{+3.02}_{-0.02}$ & $4.55^{+1.76}_{-0.49}$\\
\cline{1-2}
\multirow{ 2}{*}{PD-Ks-a-5} & a & 10:00:31.03 & +02:16:56.00 & $1.5 \pm 0.3$ & 0.5 & $5.74^{+0.03}_{-0.05}$ & $1.50^{+4.59}_{-0.01}$ & $6.89^{+0.14}_{-0.10}$\\
& b & 10:00:31.03 & +02:16:55.77 &  $2.3 \pm 0.3$ & 0.5 & $5.75^{+0.27}_{-0.05}$ & $5.65^{+0.59}_{-3.50}$ & $6.82^{+0.09}_{-0.05}$\\
\enddata
\tablecomments{Photometric redshifts derived from SED fitting the HST and JWST of the JWST-detected counterparts to the dropout candidates, using three separate routines with SED fitting codes \textsc{LePhare}, \textsc{EAZY}, and \textsc{Prospector}. We also indicate their R.A. and decl. as measured on the JWST images, the [F160W-F444W] color, and what diameter circular aperture was used to measure the photometry. The ID of each entry indicates the type of the associated IRAC-selected dropout candidate, i.e. $H$- or $K_\mathrm{s}$-band dropout. The three subsamples, i.e. secure, deboosted, and associated, are divided by horizontal lines for readability. The contaminant galaxies in the deboosted sample are omitted from this table. 
\tablenotetext{a}{The [F160W-F444W] color of this source is based on a flux upper limit in F160W so it represents a lower limit on the redness of this color. }
\tablenotetext{b}{\highlight{This source qualifies as a true dropout galaxy, such that it has no significant flux detections in the short-wavelength HST and JWST channels.}}}
\end{deluxetable*}

We consider the \textsc{Lephare} best-fit redshift valid if it satisfies the criterion 
\begin{equation}
    \frac{|z - z_{\rm LePhare}|}{1 + z_{\rm LePhare}} <0.15\,, 
\end{equation}
where $z$ corresponds to either the \textsc{EAZY} or \textsc{Prospector} photometric redshift. This is the case for the vast majority of our sample, with the exception of the following four sources: PD-H-s-3, PD-Ks-s-8, PD-Ks-d-10, and PD-Ks-a-5a. Given that their photometric redshifts and therefore the physical parameters are unreliable, we would discard them from any further analysis. However, PD-H-s-3 and PD-K-s-8 have secondary \textsc{LePhare} redshifts coinciding with the \textsc{EAZY} results, so we adopt the \textsc{LePhare} physical parameters associated with the secondary redshift solution. PD-Ks-d-10 is adjusted in a similar way, as its secondary redshift coincides with the \textsc{Prospector} solution. Lastly, PD-Ks-a-5a may have conflicting redshift results between the three SED fitting routines, but this is not the case for companion galaxy PD-Ks-a-5b, for which the \textsc{LePhare} redshift is backed up by the \textsc{EAZY} result. Given the similarity between the two sources and the large uncertainty on the \textsc{EAZY} redshift of PD-Ks-a-5a, we decide to keep the \textsc{LePhare} redshift for PD-Ks-a-5a. Therefore, none of the four JWST-detected galaxies with uncertain redshifts are discarded from the sample, but their \textsc{LePhare}-based physical parameters are modified accordingly. 

\subsubsection{General SED-derived properties of the JWST-detected galaxies}

\begin{figure}
    \plotone{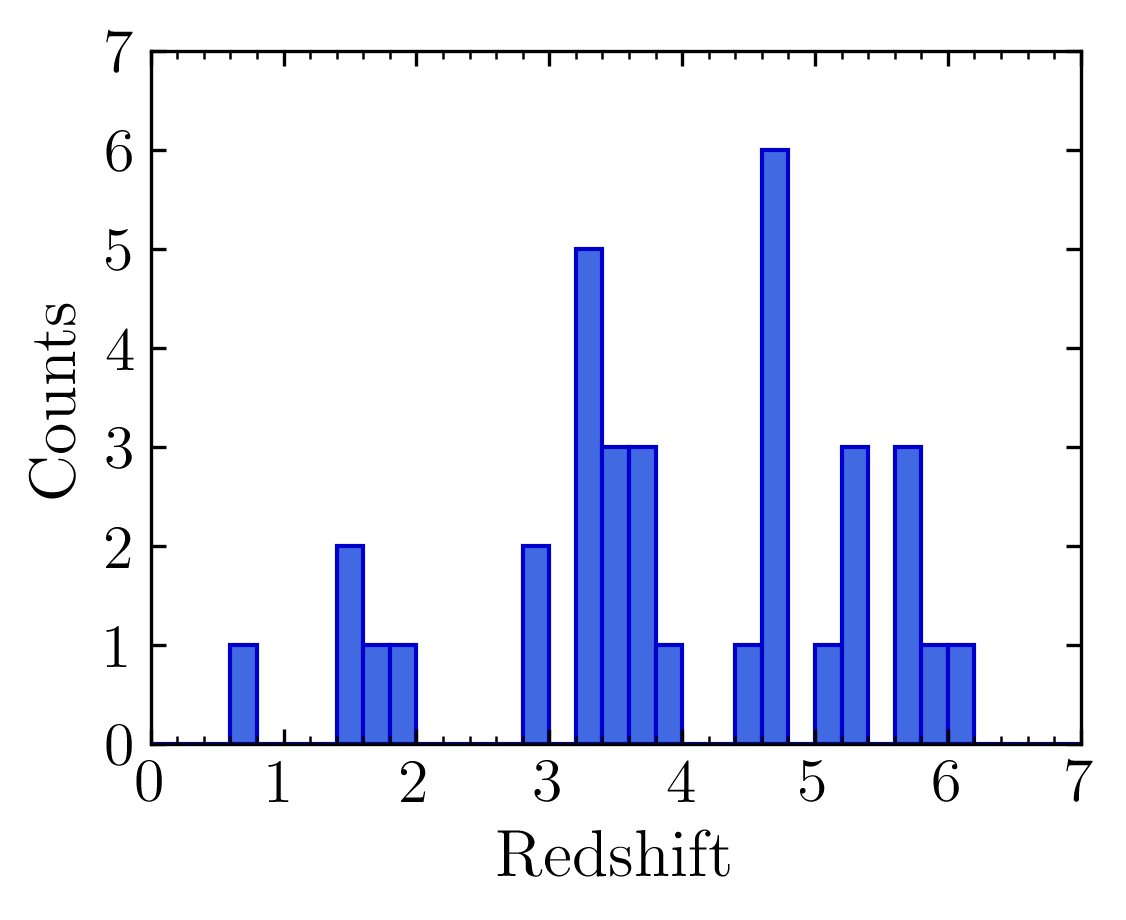}
    \caption{Photometric redshift distribution of the dropout candidates derived from the \textsc{LePhare} SED fitting the HST and JWST photometry. \label{fig:redshift}}
\end{figure}

In this section, we qualitatively analyze the physical parameters of the 35 JWST-detected counterparts to the UltraVISTA dropout candidates as derived with \textsc{LePhare}. We show the finalized \textsc{LePhare} photometric redshift distribution as discussed in the previous section in Fig.\,\ref{fig:redshift}. The vast majority of our sample is located between $z=3$ and 6. Moreover, the galaxies are relatively grouped in redshift space at $z \sim 3.3$, $z \sim 4.8$, and $z\sim 5.7$.

The physical parameters, as determined with \textsc{LePhare} are shown in Table\, \ref{tab:sedparam}, including the photometric redshift, $\chi^2_{\nu}$, color excess, stellar mass, and age. \highlight{The errors of all parameters except for the color excess represent the \nth{16} and \nth{84} percentiles. For the color excess instead, \textsc{LePhare} does not provide an error analysis for this parameter, such that we derived its uncertainty from the difference in $E(B-V)$ values between the best-fit SEDs at the lower and upper redshift error for each source. A subset of our galaxies are identified in other works, as indicated in Table\,\ref{tab:sedparam}; we discuss these individual sources in more detail in Sect.\,\ref{sect:submm}.}

\begin{deluxetable*}{lcccccccc}
 \centerwidetable 
 \tabletypesize{\footnotesize}
\tablecaption{Physical parameters of JWST-detected galaxies located within the IRAC apertures \label{tab:sedparam}}
\tablewidth{0pt}
\tablehead{
\multicolumn{2}{c}{ID} & \colhead{Redshift} & \colhead{$\chi^2_{\nu}$} &\colhead{Color Excess} & \colhead{Stellar Mass} & \colhead{Age} & \colhead{F356W} & \colhead{F444W} \\
\multicolumn{2}{c}{} & \colhead{} & \colhead{} & \colhead{$E(B-V)$} & \colhead{($\log[M/M_{\odot}]$)} & \colhead{($\times 10^{7}$ yr)} & \colhead{(AB mag)} & \colhead{(AB mag)}
}
\startdata
PD-H-s-1 && $3.28^{+0.10}_{-0.10}$ & 0.97 & $0.7 \pm 0.0$ & $10.02^{+0.14}_{-0.18}$ & $4.0^{+50.7}_{-0.0}$ & $24.09 \pm 0.05$ & $23.33 \pm 0.05$ \\
PD-H-s-2 \tablenotemark{a} && $3.32^{+0.17}_{-0.13}$ & 0.96 & $0.9 \pm 0.0$ & $9.99^{+0.13}_{-0.14}$ & $3.5^{+15.0}_{-0.7}$ & $24.07 \pm 0.05$ & $23.48 \pm 0.05$ \\
PD-H-s-3 \tablenotemark{c} && $1.61^{+0.00}_{-0.02}$ & 1.29 & $0.7 \pm 0.0$ & $8.39^{+0.29}_{-0.08}$ & $1.0^{+70.0}_{-0.0}$ & $25.22 \pm 0.05$ & $25.12 \pm 0.05$ \\
PD-Ks-s-4 && $4.71^{+0.09}_{-0.15}$ & 1.50 & $0.5 \pm 0.0$ & $9.48^{+0.10}_{-0.09}$ & $5.7^{+11.4}_{-1.2}$ & $25.14 \pm 0.05$ & $25.21 \pm 0.05$ \\
PD-Ks-s-5 && $4.78^{+0.04}_{-0.02}$ & 2.70 & $0.6 \pm 0.0$ & $10.19^{+0.09}_{-0.10}$ & $29.0^{+13.3}_{-18.6}$ & $24.24 \pm 0.05$ & $24.27 \pm 0.05$ \\
PD-Ks-s-6 \tablenotemark{b} && $4.68^{+2.03}_{-0.25}$ & 0.16 & $0.7 \pm 0.7$ & $9.66^{+0.23}_{-0.28}$ & $1.0^{+38.6}_{-0.0}$ & $24.88 \pm 0.05$ & $24.49 \pm 0.05$ \\
PD-Ks-s-7 && $4.75^{+0.06}_{-0.28}$ & 0.36 & $0.7 \pm 0.0$ & $9.98^{+0.14}_{-0.15}$ & $16.0^{+6.3}_{-12.4}$ & $24.85 \pm 0.05$ & $24.73 \pm 0.05$ \\
PD-Ks-s-8 \tablenotemark{b,c} && $3.48^{+0.23}_{-0.13}$ & 0.69 & $0.8 \pm 0.1$ & $9.90^{+0.24}_{-0.25}$ & $2.1^{+47.5}_{-0.0}$ & $24.40 \pm 0.05$ & $23.59 \pm 0.05$ \\
PD-Ks-s-9 && $4.75^{+0.06}_{-0.24}$ & 2.84 & $0.7 \pm 0.0$ & $10.10^{+0.12}_{-0.12}$ & $18.0^{+14.5}_{-12.5}$ & $24.56 \pm 0.05$ & $24.51 \pm 0.05$ \\
PD-Ks-s-10 && $5.85^{+0.39}_{-1.00}$ & 0.22 & $0.3 \pm 0.1$ & $10.12^{+0.09}_{-0.11}$ & $16.0^{+64.7}_{-0.0}$ & $25.01 \pm 0.05$ & $24.80 \pm 0.05$ \\
PD-Ks-s-11 \tablenotemark{b} && $3.51^{+0.19}_{-0.35}$ & 0.31 & $0.8 \pm 0.1$ & $9.96^{+0.25}_{-0.21}$ & $1.5^{+41.8}_{-0.0}$ & $23.92 \pm 0.05$ & $23.19 \pm 0.05$ \\[0.2cm]
\hline 
PD-H-d-1 \tablenotemark{b} \rule{0pt}{1.5\normalbaselineskip}&& $5.73^{+0.54}_{-0.14}$ & 0.88 & $0.7 \pm 0.0$ & $10.35^{+0.19}_{-0.18}$ & $1.0^{+14.8}_{-0.0}$ & $24.45 \pm 0.05$ & $23.63 \pm 0.05$ \\
PD-H-d-2 && $3.32^{+0.23}_{-0.16}$ & 0.23 & $0.5 \pm 0.0$ & $9.78^{+0.17}_{-0.24}$ & $12.8^{+62.8}_{-7.3}$ & $24.78 \pm 0.05$ & $24.31 \pm 0.05$ \\
PD-H-d-3 && $1.99^{+0.51}_{-0.09}$ & 1.23 & $0.2 \pm 0.2$ & $9.93^{+0.10}_{-0.24}$ & $300.0^{+0.0}_{-274.6}$ & $24.03 \pm 0.05$ & $23.80 \pm 0.05$ \\
PD-H-d-4 \tablenotemark{b} && $3.25^{+0.26}_{-0.07}$ & 0.88 & $1.0 \pm 0.0$ & $10.27^{+0.17}_{-0.22}$ & $3.5^{+51.4}_{-1.2}$ & $23.77 \pm 0.05$ & $23.01 \pm 0.05$ \\
PD-Ks-d-5 && $4.44^{+0.47}_{-0.30}$ & 1.3 & $0.5 \pm 0.1$ & $9.22^{+0.19}_{-0.15}$ & $1.0^{+40.2}_{-0.0}$ & $25.33 \pm 0.05$ & $25.23 \pm 0.05$ \\
PD-Ks-d-6 && $3.62^{+2.64}_{-0.16}$ & 0.49 & $0.4 \pm 0.1$ & $9.13^{+0.23}_{-0.24}$ & $2.5^{+37.7.4}_{-0.0}$ & $25.54 \pm 0.05$ & $25.15 \pm 0.05$ \\
PD-Ks-d-7 && $5.21^{+0.06}_{-0.06}$ & 0.90 & $0.2 \pm 0.0$ & $9.47^{+0.13}_{-0.13}$ & $4.0^{+33.9}_{-0.0}$ & $25.03 \pm 0.05$ & $24.69 \pm 0.05$ \\
PD-Ks-d-8 && $3.82^{+0.20}_{-1.23}$ & 0.20 & $0.1 \pm 0.3$ & $8.53^{+0.21}_{-0.26}$ & $36.0^{+22.4}_{-31.2}$ & $26.60 \pm 0.09$ & $26.63 \pm 0.13$ \\
PD-Ks-d-9 && $3.47^{+0.24}_{-0.69}$ & 0.14 & $0.6 \pm 0.1$ & $9.46^{+0.25}_{-0.22}$ & $3.0^{+48.9}_{-0.0}$ & $25.18 \pm 0.05$ & $24.66 \pm 0.05$ \\
PD-Ks-d-10 \tablenotemark{b,c} && $6.17^{+0.11}_{-0.18}$ & 0.84 & $0.4 \pm 0.0$ & $10.70^{+0.09}_{-0.11}$ & $28.6^{+33.3}_{-19.4}$ & $24.26 \pm 0.05$ & $23.89 \pm 0.05$ \\[0.2cm]
\hline 
\multirow{ 7}{*}{PD-H-a-1/2} \rule{0pt}{1.5\normalbaselineskip} & a & $2.82^{+0.48}_{-2.11}$ & 0.53 & $0.2 \pm 0.1$ & $7.28^{+0.32}_{-0.31}$ & $1.0^{+92.9}_{-0.0}$ & $28.17 \pm 0.31$ & $28.27 \pm 0.47$ \\
& b & $3.61^{+0.09}_{-0.09}$ & 2.11 & $0.3 \pm 0.1$ & $8.41^{+0.11}_{-0.08}$ & $3.5^{+7.8}_{-0.8}$ & $26.50 \pm 0.07$ & $26.11 \pm 0.06$ \\
& c & $3.61^{+1.59}_{-0.11}$ & 2.54 & $0.4 \pm 0.1$ & $8.24^{+0.10}_{-0.10}$ & $2.1^{+2.3}_{-0.6}$ & $27.25 \pm 0.13$ & $26.72 \pm 0.13$ \\
& d & $5.21^{+0.00}_{-0.02}$ & 1.22 & $0.5 \pm 0.0$ & $8.82^{+0.04}_{-0.04}$ & $1.0^{+0.5}_{-0.0}$ & $26.82 \pm 0.09$ & $25.99 \pm 0.06$ \\
& e & $5.21^{+0.00}_{-0.01}$ & 1.36 & $0.4 \pm 0.0$ & $8.89^{+0.10}_{-0.08}$ & $1.0^{+2.1}_{-0.0}$ & $26.11 \pm 0.05$ & $25.50 \pm 0.05$ \\
& f & $5.18^{+0.04}_{-0.01}$ & 1.16 & $0.5 \pm 0.0$ & $9.65^{+0.28}_{-0.14}$ & $1.0^{+29.5}_{-0.0}$ & $24.50 \pm 0.05$ & $23.98 \pm 0.05$ \\ 
& g & $0.77^{+0.11}_{-0.13}$ & 0.12 & $0.1 \pm 0.1$ & $7.39^{+0.17}_{-0.22}$ & $11.4^{+150.3}_{-5.5}$ & $26.84 \pm 0.09$ & $27.10 \pm 0.19$ \\
\cline{1-2}
\multirow{ 2}{*}{PD-Ks-a-3} & a & $1.52^{+0.10}_{-0.13}$ & 0.63 & $0.1 \pm 0.1$ & $7.09^{+0.16}_{-0.15}$ & $1.5^{+20.6}_{-0.0}$ & $27.37 \pm 0.11$ & $27.08 \pm 0.12$ \\
& b & $1.57^{+0.05}_{-0.29}$ & 0.80 & $0.3 \pm 0.3$ & $7.58^{+0.16}_{-0.15}$ & $1.0^{+60.6}_{-0.0}$ & $26.91 \pm 0.07$ & $26.86 \pm 0.09$ \\
\cline{1-2}
\multirow{ 3}{*}{PD-Ks-a-4} & a & $2.86^{+0.33}_{-1.46}$ & 0.47 & $0.0 \pm 0.2$ & $8.47^{+0.27}_{-0.33}$ & $200.0^{+0.0}_{-196.1}$ & $26.78 \pm 0.13$ & $26.58 \pm 0.17$ \\
& b & $3.36^{+1.81}_{-2.30}$ & 0.17 & $0.0 \pm 0.0$ & $7.39^{+0.16}_{-0.15}$ & $3.0^{+13.3}_{-0.6}$ & $28.32 \pm 0.53$ & $27.64 \pm 0.42$ \\
& c & $4.76^{+0.09}_{-0.77}$ & 0.94 & $0.6 \pm 0.4$ & $9.19^{+0.29}_{-0.23}$ & $1.0^{+34.3}_{-0.0}$ & $25.74 \pm 0.05$ & $25.51 \pm 0.07$ \\
\cline{1-2}
\multirow{ 2}{*}{PD-Ks-a-5} & a & $5.74^{+0.03}_{-0.05}$ & 1.66 & $0.2 \pm 0.0$ & $9.13^{+0.19}_{-0.18}$ & $3.5^{+55.3}_{-0.1}$ & $26.25 \pm 0.05$ & $25.31 \pm 0.05$ \\
& b & $5.75^{+0.27}_{-0.05}$ & 3.84 & $0.5 \pm 0.0$ & $9.41^{+0.19}_{-0.17}$ & $1.0^{+33.0}_{-0.0}$  & $25.82 \pm 0.05$ & $24.81 \pm 0.05$ \\
\enddata
\tablecomments{Physical parameters derived from SED fitting the HST and JWST of the JWST-detected counterparts to the UltraVISTA dropout candidates, using \textsc{LePhare}. The contaminant galaxies in the deboosted sample are omitted from this table. 
\tablenotetext{a}{Detected at X-ray wavelengths in Chandra/XMM-Newton \citep{cappelluti2009,civano2016}}
\tablenotetext{b}{Detected at submillimeter wavelengths with ALMA/SCUBA/VLA \citep{koprowski2016,smolcic2017,liu2019,simpson2019,simpson2020}}
\tablenotetext{c}{The redshift and physical parameters presented for this source are based on the second best-fit model from \textsc{LePhare}, which does agree with the redshift results from \textsc{EAZY} and/or \textsc{Prospector}. }}
\end{deluxetable*}

\begin{figure}
    \plotone{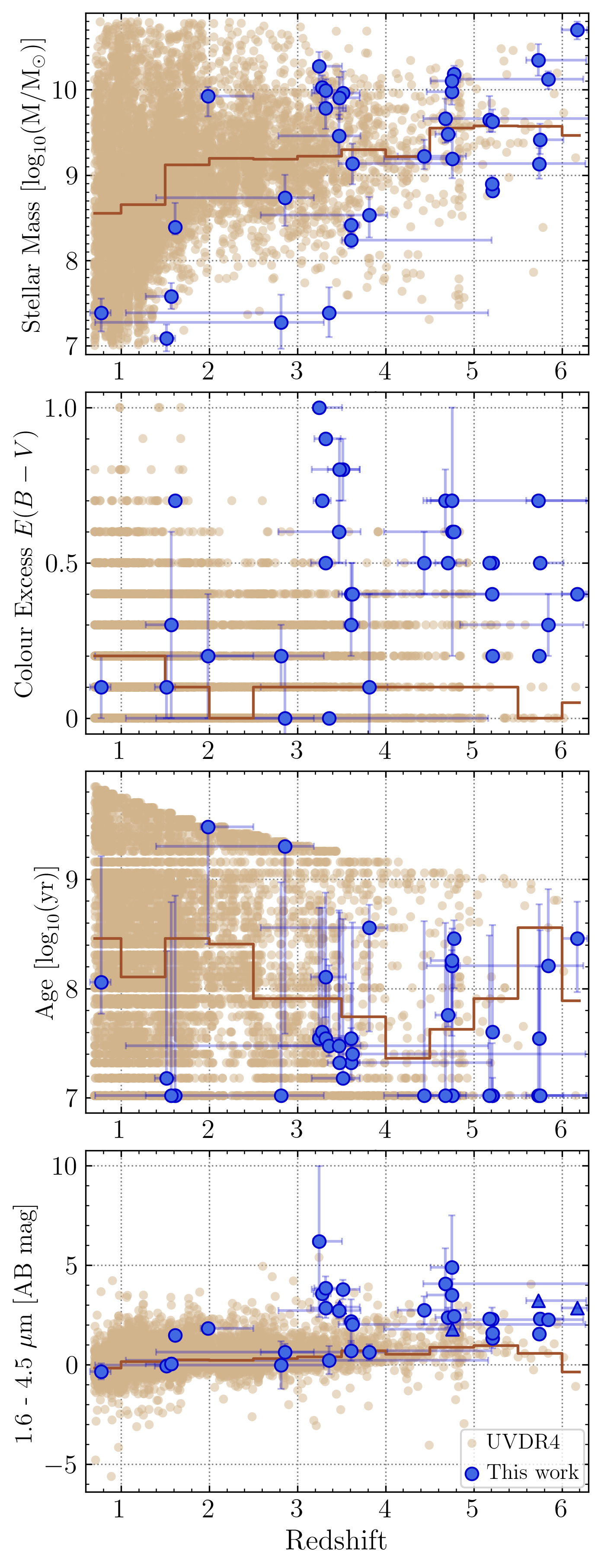}
    \caption{Stellar mass, color excess, age, and 1.6--4.5 $\mu$m color against redshift for our JWST-detected sample in blue. Ancillary UltraVISTA DR4 galaxies within the same redshift and mass range as the dropout candidates are shown in tan data points, with their median value derived in $dz=0.5$ bins indicated by the brown line. The 1.6--4.5 $\mu$m color for the dropout candidates is based on the WFC3/F160W and NIRCam/F444W bands, and on the VIRCAM $H$ and IRAC 4.5 $\mu$m bands for the ancillary sample. Upper and lower limits on the redness of this color are indicated with downward- and upward-pointing triangles respectively. \label{fig:sedparam}}
\end{figure}

Fig.\,\ref{fig:sedparam} shows the stellar mass, color excess, age, and observed [F160W-F444W] color against redshift for our JWST-detected sample. To put these special galaxies into the context of the general population, we compare them to the UltraVISTA DR4 catalog from \citet{mierlo2022}. This catalog contains physical parameters for UltraVISTA $HK_{\rm s}$-detected galaxies at $z_{\rm phot}=0$--9, derived from SED fitting on multiwavelength photometry spanning optical to IRAC imaging. To ensure a fair comparison, the catalog has been cut to the PRIMER area, such that it contains \highlight{10,272 galaxies}. The UltraVISTA DR4 galaxies shown in Fig.\,\ref{fig:sedparam} are matched to the redshift and stellar mass range of the JWST-detected sample, i.e., \highlight{$z=0.7$--6.2 and $\log(M_*/M_\odot)=7.0$--10.8.}

We find that \highlight{$\sim70$\,\%} of our JWST-detected sample is defined as $z=3$--6 galaxies of intermediate stellar mass $\log(M_*/M_\odot) \sim 8.2$--10.3, which have red [F160W-F444W] colors explained by their dusty nature. In fact, for the secure and deboosted samples (where there is good correspondence between the IRAC- and JWST-detected sources), \highlight{95\%} of the galaxies are dusty with $E(B-V)\geq 0.2$, and moreover, \highlight{71\%} are highly dusty with $E(B-V)\geq 0.5$. Their corresponding red colors explain straightforwardly why these sources were initially identified as dropouts in UltraVISTA imaging.

Although our sample consists mostly of these dusty intermediate-$z$ sources, the remaining JWST-detected galaxies display a variety in of behaviors. \highlight{For example, source PD-H-d-3, with $z=1.99$ and $E(B-V)=0.2$, appears to be an $H$-band dropout galaxy in the UltraVISTA imaging. The HST and JWST photometry reveal that PD-H-d-3 is in fact not a dropout galaxy, but that its red color can be attributed to its age of 3.0 Gyr.} In addition, system PD-Ks-a-3 is very different from any of the other dropout candidates we study throughout this paper \highlight{(see Fig.\,\ref{fig:stamps_associated} for the image cutouts)}. It is in fact the faintest IRAC source in our sample, with magnitude 27.0 ($3.5\sigma$) in the IRAC 4.5 $\mu$m band and no detection at all in the 3.6 $\mu$m band. HST and JWST imaging reveal that this IRAC source is made up of two young galaxies at $z\sim1.5$, that have very flat SEDs and low stellar masses of $\log(M_*/M_\odot)=7.1$ and 7.6. To understand why this duo enters our dropout candidate sample, we search for galaxies of comparable redshift and IRAC brightness in the UltraVISTA DR4 catalog. This yields only six galaxies, all of which have 3--$22\sigma$ detections in the VIRCAM $H$, $K_\mathrm{s}$ and IRAC 3.6 $\mu$m bands. Given that the IRAC and VIRCAM imaging considered in this work are of comparable depths, it is unclear why PD-Ks-a-3-a and -b are absent from the VIRCAM imaging. Finally, we note the arc-like spatial alignment as seen in \textbf{PD-H-a-1/2}, which are two IRAC-detected galaxies at $1\farcs4$ angular separation that drop out in the VIRCAM $H$ band. However, the HST and JWST imaging reveal that these two IRAC sources are in fact made up of seven distinct galaxies that are partially spatially aligned, two of them at $z=3.6$ and three at $z=5.2$, although PD-H-a-1/2-c has a significant secondary solution at $z=5.2$ too. 

In the context of the general galaxy population, Fig.\,\ref{fig:sedparam} clearly shows that the JWST-detected galaxies in this work are amongst the reddest and dustiest intermediate-mass sources at $z\simeq3-6$. However, there are several UltraVISTA DR4 galaxies with similarly red colors, i.e. $[H-4.5]\gtrsim 2$, which do not drop out of the VIRCAM images, despite being of comparable brightness in the IRAC bands as our JWST-detected sample. Upon further inspection, these UltraVISTA DR4 galaxies are mostly old galaxies with ages exceeding 1 Gyr, such that given their redshift, the well-developed Balmer break is located in between the $H$ and $K_\mathrm{s}$ bands, explaining their red $[H-4.5]$ color and simultaneous presence in the UltraVISTA DR4 catalog (which was detected on the stacked $HK_\mathrm{s}$ image). In addition, a small sample of the red UltraVISTA DR4 galaxies are dusty with $E(B-V)=0.3$--0.7, but overall brighter than our JWST-detected galaxies ($\sim 1.2$ mag in the IRAC bands), such that they are just bright enough to be detected in the UltraVISTA imaging. 

\begin{figure}
    \plotone{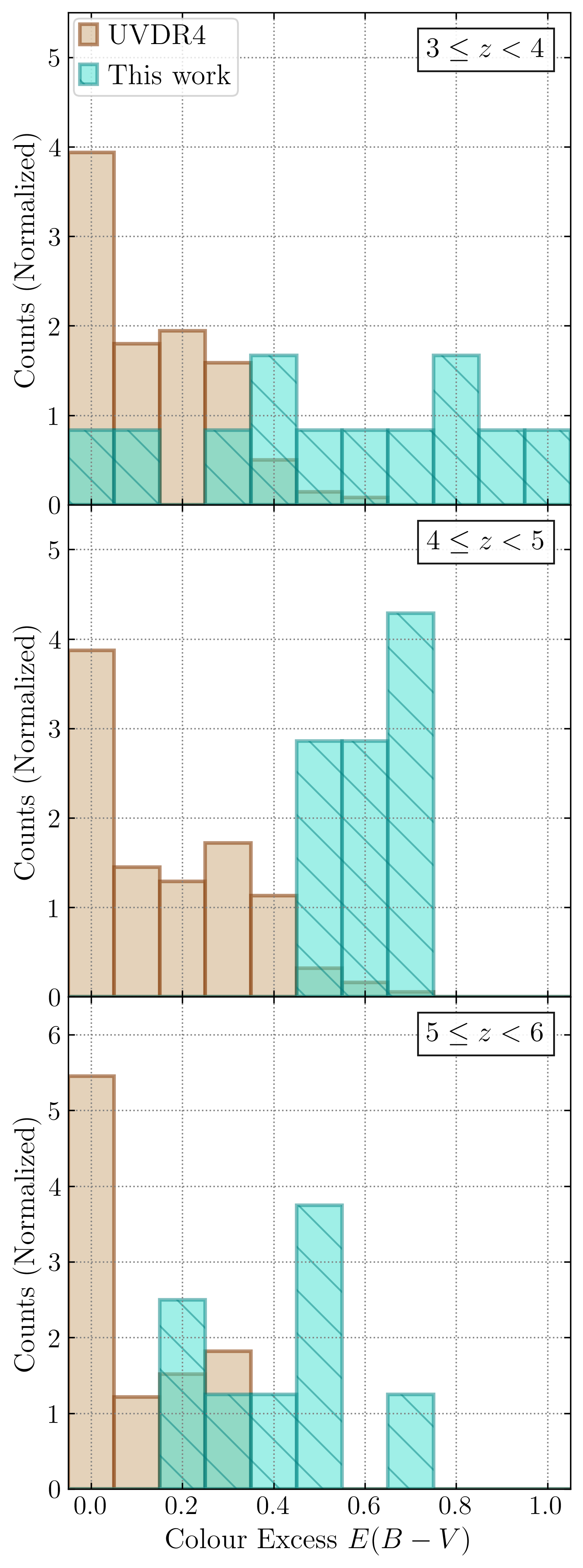}
    \caption{Normalized color excess distribution of the JWST-detected galaxies in this work (turquoise) and the UltraVISTA DR4 galaxies (tan), in three redshift bins $z=3$--4, $z=4$--5, and $z=5$--6. \label{fig:ebv_hist}}
\end{figure}

We show the normalized color excess distributions of our JWST-detected galaxies and the UltraVISTA DR4 galaxies in three redshift bins between $z=3$ and 6 in Fig.\,\ref{fig:ebv_hist}. 
In each redshift bin, we find that the color excess distribution between the UltraVISTA DR4 galaxies and our sample is distinctly different, as confirmed by performing the two-sample Kolmogorov-Smirnov test \citep{smirnov1939}, which yields $p$-values \highlight{$<3 \times 10^{-3}$}. The JWST-detected galaxies are consistently significantly dustier, even though \citet{deshmukh2018} already showed that the fraction of $E(B-V)\geq 0.2$ galaxies at the epoch $z=3$--6 is considered at $\sim30$--40\%. Especially at $z=4$--5, \highlight{all of the JWST-detected galaxies} are highly dusty, with $E(B-V)\geq 0.5$. \highlight{Our results imply these dust-reddened galaxies are a special population of galaxies that were not characterized by previous wide-area near-infrared surveys.} 

\subsection{H$\alpha$ emission in the UltraVISTA dropout galaxies} \label{sect:halpha}

From inspection of the \textsc{LePhare} best-fit SEDs, some of the JWST-detected galaxies appear to have excess flux in bands coinciding with the (H$\alpha$+ N[II] + S[II]) line complex considering their photometric redshifts. Following \citet{caputi2017}, we set a minimum condition of $\mathrm{\Delta mag = mag_{\rm obs} - mag_{\rm synth} < -0.1}$ in the band to consider the flux excess significant. Here, $\mathrm{mag_{\rm synth}}$ is the continuum emission in the band, which we obtain by remodelling the SEDs with \textsc{LePhare} whilst excluding the affected band and emission-line modelling, and adopting the fixed photometric redshift obtained from the original \textsc{LePhare} run (with all bands and emission lines). In addition, to ensure that flux excess is not the result of a poorly constrained continuum, we imposed the condition that $\mathrm{\Delta mag \leq 2 \times mag_{\rm err}}$ in at least one adjacent observed band. 

\highlight{For all galaxies where the H$\alpha$ line coincides with one of the HST and JWST bands}, we calculated the rest-frame equivalent width ($\mathrm{EW_0}$) of the (H$\alpha$+ N[II] + S[II]) line following the approach outlined in \citep{rinaldi2023} (which is in turn based on \citealt{marmol2016}), such that \begin{equation}
    \mathrm{EW_0 = \frac{W_{rec}}{1+z}(10^{-0.4\Delta mag} -1),}
\end{equation}
where $\mathrm{W_{rec}}$ is the FWHM of the filter containing the emission line of interest. We propagate the redshift error and the flux error into the $\mathrm{EW_0}$ error. Subsequently, we converted the flux excess into an H$\alpha$ emission-line flux using the modeled continuum flux, \highlight{such that 
\begin{equation}
    f(\mathrm{H\alpha+N[II]+S[II]}) = 10^{-0.4\Delta \mathrm{mag}} \times f_{\mathrm{obs}} - f_{\mathrm{synth}},
\end{equation}
where $f_{\mathrm{obs}}$ and $f_{\mathrm{synth}}$ are the observed and modeled fluxes, respectively, in the filter containing the emission line, where we again propagate the flux error.} We corrected for dust extinction using the color excess from the original SED fitting and assuming a \citet{calzetti2000} dust reddening law. Here, we assumed that the extinction is the same for both the lines and the continuum. We also propagated the error on the dust extinction, which was derived from the difference in $E(B-V)$ values between the best-fit SEDs at the lower and upper redshift error for each source. We obtained the net H$\alpha$ flux considering that $f$(H$\alpha$)=0.63$f$(H$\alpha$+N[II]+S[II]) \citep{anders2003}, and converted it to a SFR following \citet{kennicutt1998}:
\begin{equation}
    \mathrm{SFR(H\alpha)} (M_\odot\, \mathrm{yr^{-1}) = 7.936 \times 10^{-42}} \times L_\mathrm{{H\alpha}} \mathrm{(erg\, s^{-1}),} 
\end{equation}
where we divide the resulting SFRs by a factor of 1.69 to re-scale them from a \citet{salpeter1955} IMF over (0.1--100) $M_\odot$ to a \citet{chabrier2003} IMF. We calculate the specific SFR (sSFR) by dividing the SFR by the stellar mass. We propagate the uncertainties on the SFR and stellar mass into a sSFR uncertainty. \highlight{However, propagating the asymmetric uncertainty on the stellar mass is not trivial, such that we determine an average error on the stellar mass from its uncertainty limits and propagate that instead. Therefore, we warn the reader that the reported values on the sSFR uncertainty are only approximate.}

Not all galaxies in our JWST-detected sample display flux excess at the location of the H$\alpha$ line complex; for those that do not, we calculated upper limits on their H$\alpha$ properties from the minimum line complex $\mathrm{EW_0}$ that would produce $\mathrm{\Delta mag = mag_{obs} - mag_{synth} = -0.1}$ in the band containing the emission-line complex. \highlight{Only for source PD-H-s-3 are we unable to calculate information on the H$\alpha$ emission; due to its redshift, the line does not fall into any of the HST and JWST filters considered in this study.} 

\begin{deluxetable*}{llcccccc}
 \centerwidetable 
 \tabletypesize{\footnotesize}
\tablecaption{H$\alpha$ $\mathrm{EW_0}$, $L$ and SFR of JWST-detected galaxies located within the IRAC apertures, including upper limits for sources that display no flux excess. \label{tab:halpha}}
\tablewidth{0pt}
\tablehead{
\multicolumn{2}{c}{ID} & \colhead{Redshift} & \colhead{Band} & \colhead{ (H$\alpha$+ N[II] + S[II]) $\mathrm{EW_0}$} & \colhead{$L_\mathrm{{H \alpha}}$} & \colhead{SFR} & \colhead{sSFR} \\
\multicolumn{2}{c}{} & \colhead{} & \colhead{} & \colhead{(\AA)} & \colhead{($10^{42}\,\mathrm{erg\, s^{-1}}$)} & \colhead{($M_\odot \mathrm{yr^{-1}}$)} & \colhead{($\log(\mathrm{yr^{-1}})$})}
\startdata
PD-H-s-1 & & 3.28 & F277W & $326^{+15}_{-15}$ & $3.2 \pm 0.1$ & $15 \pm 6$ & $-8.8 \pm 0.2$\\
PD-H-s-2 \tablenotemark{a} & & 3.32 & F277W & $722^{+34}_{-32}$ & $3.8 \pm 0.2$ & - & -\\
PD-H-s-3 \tablenotemark{d} & & 1.61 & - & - & - & - & -\\
PD-Ks-s-4 & & 4.71 & F356W & $667^{+32}_{-30}$ & $13.9 \pm 0.6$ & $65 \pm 27$ & $-7.7 \pm 0.2$\\
PD-Ks-s-5 & & 4.78 & F356W & $617^{+29}_{-28}$ & $23.2 \pm 1.1$ & $109 \pm 46$ & $-8.1 \pm 0.2$\\
PD-Ks-s-6 & & 4.68 & F356W & $622^{+30}_{-21}$ & $9.0 \pm 2.8$ & $42 \pm 25$ & $-8.0 \pm 0.4$\\
PD-Ks-s-7 & & 4.75 & F356W & $594^{+29}_{-27}$ & $9.4 \pm 0.4$ & $44 \pm 18$ & $-8.3 \pm 0.2$\\
PD-Ks-s-8 & & 3.48 & F277W & $395^{+19}_{-17}$ & $2.0 \pm 0.6$ & $9 \pm 5$ & $-8.9 \pm 0.4$\\
PD-Ks-s-9 & & 4.75 & F356W & $538^{+26}_{-25}$ & $11.4 \pm 0.5$ & $54 \pm 22$ & $-8.4 \pm 0.2$\\
PD-Ks-s-10 \tablenotemark{b} & & 5.85 & F444W & $155$ & $18.6$ & $87$ & -8.2\\
PD-Ks-s-11 & & 3.51 & F277W & $653^{+33}_{-29}$ & $4.7 \pm 1.5$ & $22 \pm 13$ & $-8.6 \pm 0.3$ \\[0.2cm]
\hline 
PD-H-d-1 \rule{0pt}{1.5\normalbaselineskip} & & 5.73 & F444W & $400^{+19}_{-17}$ & $30.4 \pm 1.4$ & $143 \pm 59$ & $-8.2 \pm 0.3$\\
PD-H-d-2 \tablenotemark{b} & & 3.32 & F277W & $158$ & $1.7$ & $8$ & -8.9\\
PD-H-d-3 \tablenotemark{b} & & 1.99 & F200W & $229$ & $2.7$ & $13$ & -8.8\\
PD-H-d-4 & & 3.25 & F277W & $437^{+20}_{-19}$ & $1.8 \pm 0.1$ & $8 \pm 3$ & $-9.4 \pm 0.3$\\
PD-Ks-d-5 & & 4.44 & F356W & $358^{+17}_{-15}$ & $6.2 \pm 1.9$ & $29 \pm 17$ & $-7.8 \pm 0.3$\\
PD-Ks-d-6 & & 3.62 & F277W & $311^{+17}_{-11}$ & $3.1 \pm 0.9$ & $14 \pm 9$ & $-8.0 \pm 0.4$\\
PD-Ks-d-7 & & 5.23 & F410M & $484^{+26}_{-25}$ & $74.2 \pm 3.9$ & $349 \pm 146$ & $-6.9 \pm 0.2$\\
PD-Ks-d-8 \tablenotemark{b} & & 3.82 & F356W & $169$ & $2.1$ & $10$ & -7.5\\
PD-Ks-d-9 \tablenotemark{b} & & 3.47 & F277W & $153$ & $0.9$ & $4$ & -8.8\\
PD-Ks-d-10 & & 6.17 & F444W & $318^{+15}_{-14}$ & $62.2 \pm 2.9$ & $292 \pm 122$ & $-8.2 \pm 0.2$\\[0.2cm]
\hline 
\multirow{ 7}{*}{PD-H-a-1/2} \rule{0pt}{1.5\normalbaselineskip} & a & 2.82 & F277W & $1602^{+675}_{-268}$ & $1.4 \pm 0.5$ & $7 \pm 4$ & $-6.5 \pm 0.4$\\
& b & 3.61 & F277W & $1091^{+51}_{-49}$ & $7.7 \pm 2.4$ & $36 \pm 21$ & $-6.9 \pm 0.3$\\
& c \tablenotemark{c} & 3.61 & F277W & $2169^{+147}_{-105}$ & $5.4 \pm 1.7$ & $26 \pm 15$ & $-6.8 \pm 0.3$\\
& d & 5.21 & F410M & $1171^{+64}_{-64}$ & $23.9 \pm 1.3$ & $112 \pm 47$ & $-6.8 \pm 0.2$\\
& e & 5.21 & F410M & $811^{+37}_{-37}$ & $39.6 \pm 1.8$ & $186 \pm 77$ & $-6.6 \pm 0.2$\\
& f & 5.18 & F410M & $515^{+24}_{-24}$ & $79.8 \pm 3.7$ & $375 \pm 156$ & $-7.1 \pm 0.3$\\
& g \tablenotemark{b} & 0.77 & F115W & $460$ & $0.04$ & $0.2$ & -8.1\\
\cline{1-2}
\multirow{ 2}{*}{PD-Ks-a-3} & a & $1.52$ & F160W & $115$ & $0.2$ & $0.8$ & -7.2\\
& b \tablenotemark{b} & $1.57$ & F160W & $113$ & $0.1$ & 0.6 & -7.8\\
\cline{1-2}
\multirow{ 3}{*}{PD-Ks-a-4} & a & $2.86$ & F277W & $355^{+72}_{-41}$ & $2.4 \pm 1.5$ & $11 \pm 12$ & $-7.7 \pm 0.5$\\
& b \tablenotemark{b}& $3.36$ & F277W & $157$ & $0.6$ & $3$ & -7.0\\
& c & $4.76$ & F356W & $446^{+24}_{-20}$ & $4.5 \pm 5.6$ & $21 \pm 55$ & $-7.9 \pm 1.1$\\
\cline{1-2}
\multirow{ 2}{*}{PD-Ks-a-5} & a & $5.74$ & F444W & $1641^{+76}_{-75}$ & $76.9 \pm 3.5$ & $361 \pm 151$ & $-6.6 \pm 0.3$\\
& b & $5.75$ & F444W & $1577^{+73}_{-70}$ & $50.0 \pm 2.2$ & $225 \pm 94$ & $-7.1 \pm 0.3$\\
\enddata
\tablecomments{Measurements for the (H$\alpha$+ N[II] + S[II]) rest-frame equivalent width, the H$\alpha$ luminosity and the derived (specific) star formation rate for the JWST-detected counterparts to our UltraVISTA dropout candidates. 
\tablenotetext{a}{This source is likely an AGN because of ancillary X-ray emission, such that the SFR conversion from \citet{kennicutt1998} is not applicable. }
\tablenotetext{b}{No flux excess in the band where the H$\alpha$ line should be observed for this source. The $EW_{0}$, $L_\mathrm{{H\alpha}}$ and SFR are upper limits assuming $\mathrm{\Delta mag = 0.1}$.}
\tablenotetext{c}{This source has poorly constrained continuum in bands F356W and F410M, and a significant secondary solution at $z=5.2$, such that the H$\alpha$ properties listed are uncertain at best.}
\tablenotetext{d}{Due to its redshift, the H$\alpha$ emission line falls outside any of the bands considered in this work, such that estimating its flux excess is not possible.}}
\end{deluxetable*}

Table\,\ref{tab:halpha} lists the (H$\alpha$+ N[II] + S[II]) $\mathrm{EW_0}$, the H$\alpha$ luminosity, and the derived SFR for each JWST-detected counterpart to the UltraVISTA dropout candidates, including the upper limits for the sources which show no flux excess. We also list the band in which the flux excess is observed and the sSFR. For this and any further plots involving the stellar mass, we consider the stellar mass as reported in Table\,\ref{fig:sedparam}, i.e., derived from the original SED fitting with emission lines and the filter containing the H$\alpha$ line. In total, we identify H$\alpha$ emitters from the HST and JWST photometry amongst \highlight{19 of the 26} IRAC-selected dropout candidates, such that \highlight{$\sim 75$\,\%} of the sources that enter our original dropout candidate selection are boosted redward of $K_{\mathrm{s}}$ band due to the presence of H$\alpha$ line complex emission. 

\begin{figure}
    \plotone{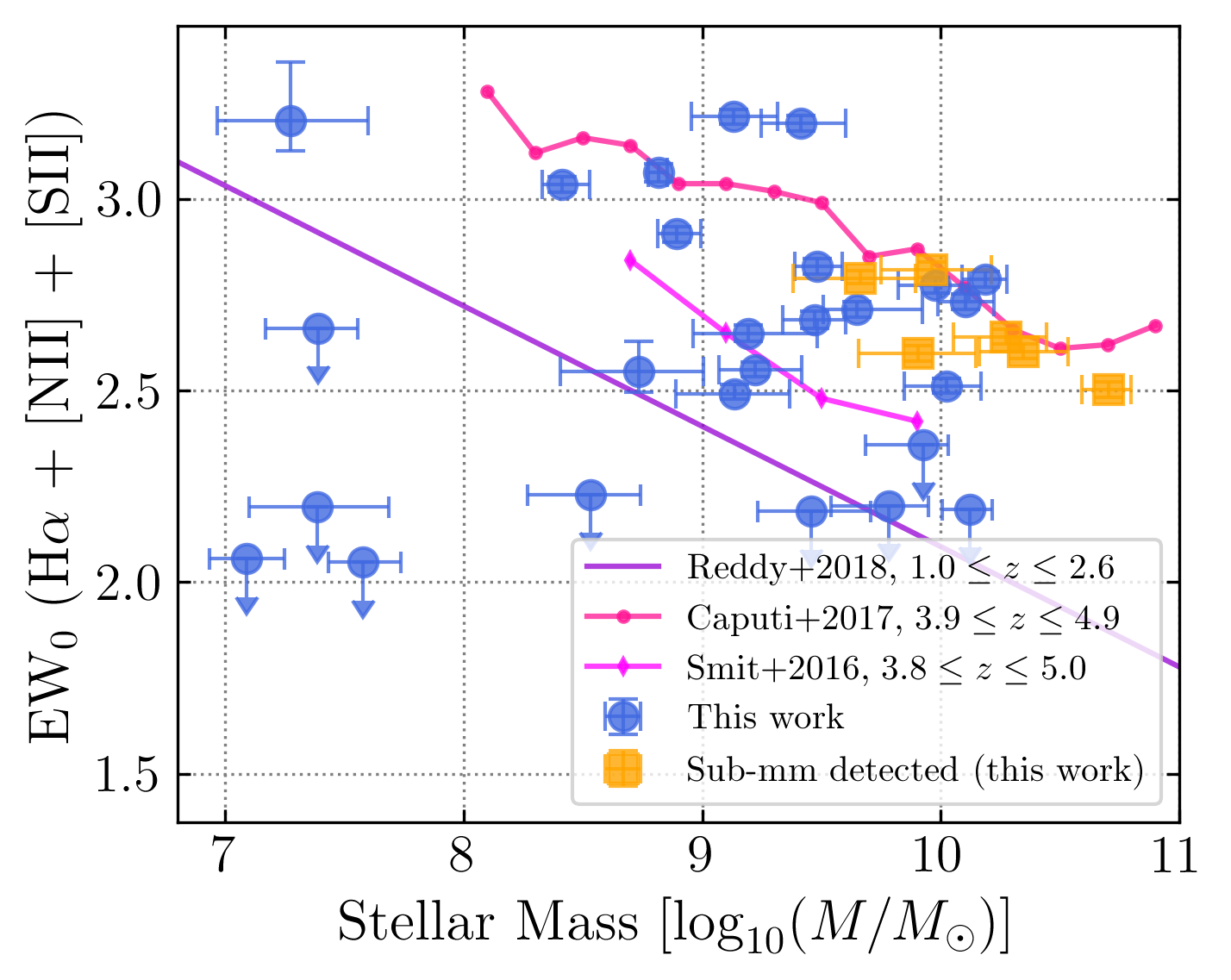}
    \caption{ (H$\alpha$+ N[II] + S[II]) rest-frame equivalent width vs. stellar mass for all JWST-detected galaxies in our JWST-detected sample. $\mathrm{EW_{0}}$ upper limits are indicated with downward-pointing arrows. The orange data points represent the six submillimeter-detected galaxies in the sample. We also show the median (H$\alpha$+ N[II] + S[II]) $\mathrm{EW_{0}}$ results from \citet{smit2016} (photometric--$z$ sample) and \citet{caputi2017}, from galaxies at $z=3.8$--5.0 and $z=3.9$--4.9, in pink circles and magenta diamonds, respectively. The purple line shows the empirical relation between (H$\alpha$+ N[II]) and $M_*$ from \citet{reddy2018}, derived from spectroscopic line measurements for galaxies at $z=1.0$--2.6. \label{fig:halpha_ew}}
\end{figure}

Fig.\,\ref{fig:halpha_ew} shows the derived (H$\alpha$+ N[II] + S[II]) $\mathrm{EW_0}$ versus stellar mass for our JWST-detected sample. We exclude here and from any further plots the sources PD-H-s-2 and PD-H-a-1/2 c; the former because its detection at X-ray wavelengths means it is most likely an active galactic nucleus (AGN), such that the SFR conversion from \citet{kennicutt1998} cannot be applied. The latter we exclude as both its redshift and the continuum in the bands redward to the H$\alpha$ line complex are poorly constrained, such that its listed H$\alpha$ properties in Table\,\ref{tab:halpha} are uncertain at best. We highlight in our plots six sources that are detected at sub/millimeter wavelengths and are all H$\alpha$ emitters, which we discuss further in Sect.\,\ref{sect:submm}. The rest-frame EWs span values between \highlight{$\sim 100$--2200\,\AA.} We find a general decline in the median $\mathrm{EW_0}$ values with stellar mass: they are $\sim 950$\,\AA\ for galaxies with $\log(M_*/M_\odot)=8.0$--9.0,\highlight{but only $\sim 440$\,\AA\ for galaxies with $\log(M_*/M_\odot)=10.0$--10.5}. \highlight{The anticorrelation between the main optical emission lines and stellar mass is well-known at different redshifts (e.g, \citealt{reddy2018,endsley2021,caputi2023,rinaldi2023}).} This trend is shown in Fig.\,\ref{fig:halpha_ew}, including the median (H$\alpha$+ N[II] + S[II]) $\mathrm{EW_{0}}$ measurements from \citet{smit2016} (the photometric--$z$ sample) and \citet{caputi2017}, for galaxies at $z=3.8$--5.0 and $z=3.9$--4.9 respectively. We also show the empirical relation between (H$\alpha$+ N[II]) $\rm EW_0$ and $M_*$ from \citet{reddy2018}, derived from spectroscopic line measurements for galaxies at $z=1.0$--2.6. The difference in $\rm EW_0$ between the (H$\alpha$+ N[II]+S[II]) and (H$\alpha$+ N[II]) line complex is on average only 0.09 dex, such that we can safely include their result in our plot. The $\mathrm{EW_0}$ measurements in this work are located above this empirical relation, as expected given that emission-line equivalent widths increase with the redshift at a fixed stellar mass \citep{reddy2018}. We attribute the difference between the results from \citet{smit2016} and \citet{caputi2017} to cosmic variance and \highlight{a different level of statistics}, as the latter study covers a field that is $\sim 7$ times larger than that of the former. Finally, we note that the observed negative $\rm EW_{0}$-$M_*$ correlation for our JWST-detected galaxies makes sense from the perspective that these are IRAC-selected sources with no UltraVISTA $H$- and/or $K_{\rm s}$-band counterparts; the less massive galaxies require significant flux boosting from H$\alpha$ emission to enter our sample. 

\begin{figure}
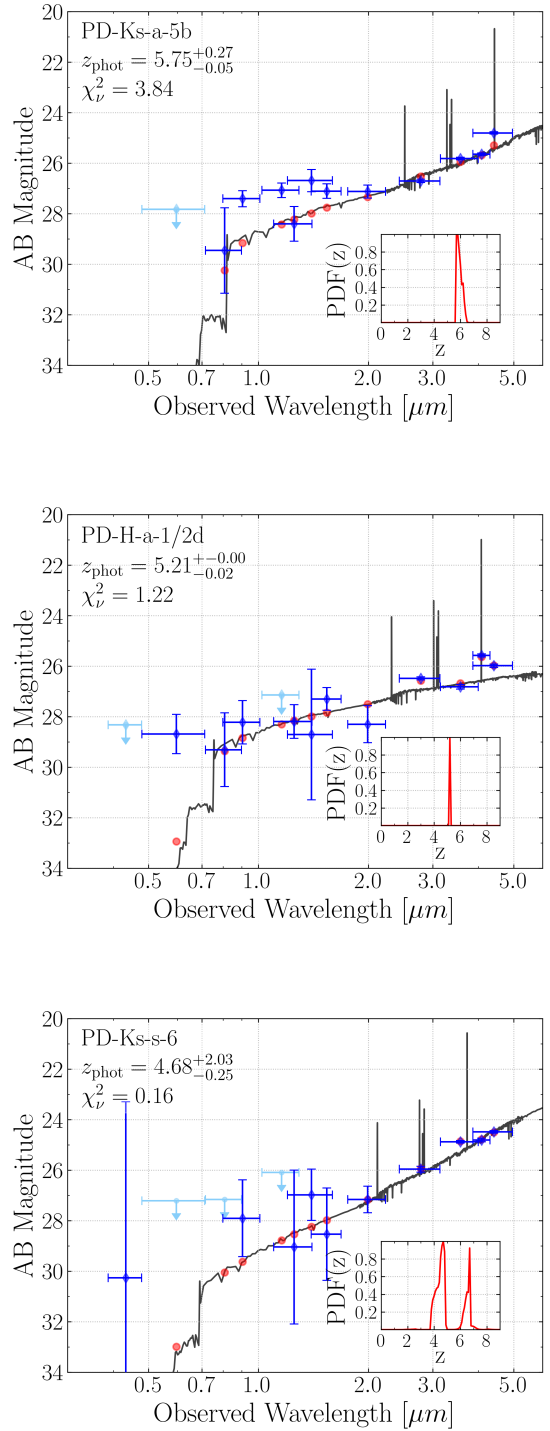

    \gridline{\fig{seds/plot_SED_8716.png}{0.4\textwidth}{}}
    \gridline{\fig{seds/plot_SED_9946.png}{0.4\textwidth}{}}
    \gridline{\fig{seds/plot_SED_31541.png}{0.4\textwidth}{}}
    \caption{Best-fit SED obtained with \textsc{LePhare} on HST and JWST photometry for three of the most prominent H$\alpha$ emitters among the JWST-detected galaxies. Photometric measurements are shown in blue diamonds, flux upper limits in cyan with downward-pointing arrows, and template fluxes (with emission line contributions) with red circles. \label{fig:sed_halpha}}
\end{figure}

We show the best-fit SEDs (derived from the original \textsc{LePhare} run) for \highlight{three} of the most prominent H$\alpha$ emitters in our sample in Fig.\,\ref{fig:sed_halpha}. We show the SED for PD-Ks-a-5-b, which together with its associated galaxy are amongst the highest redshift galaxies in our sample that display flux excesses, with $z=5.75$ and $\mathrm{EW_{0}}\sim 1600$\,\AA. In addition, we show the SED of PD-H-a-1/2-d, which is the brightest in this system of sources, where all but the one contaminating foreground galaxy are strong H$\alpha$ emitters, with mean $\mathrm{EW_{0}}\sim 1040$\,\AA. Lastly, we include the SED of one of the sources detected at submillimeter wavelengths, i.e. PD-Ks-s-6, \highlight{which due to extreme dust content combined with its H$\alpha$ excess, has a red slope perfectly demonstrating why this relatively IRAC-bright galaxy is undetected in the ground-based VIRCAM imaging.} 

\begin{figure}
    \plotone{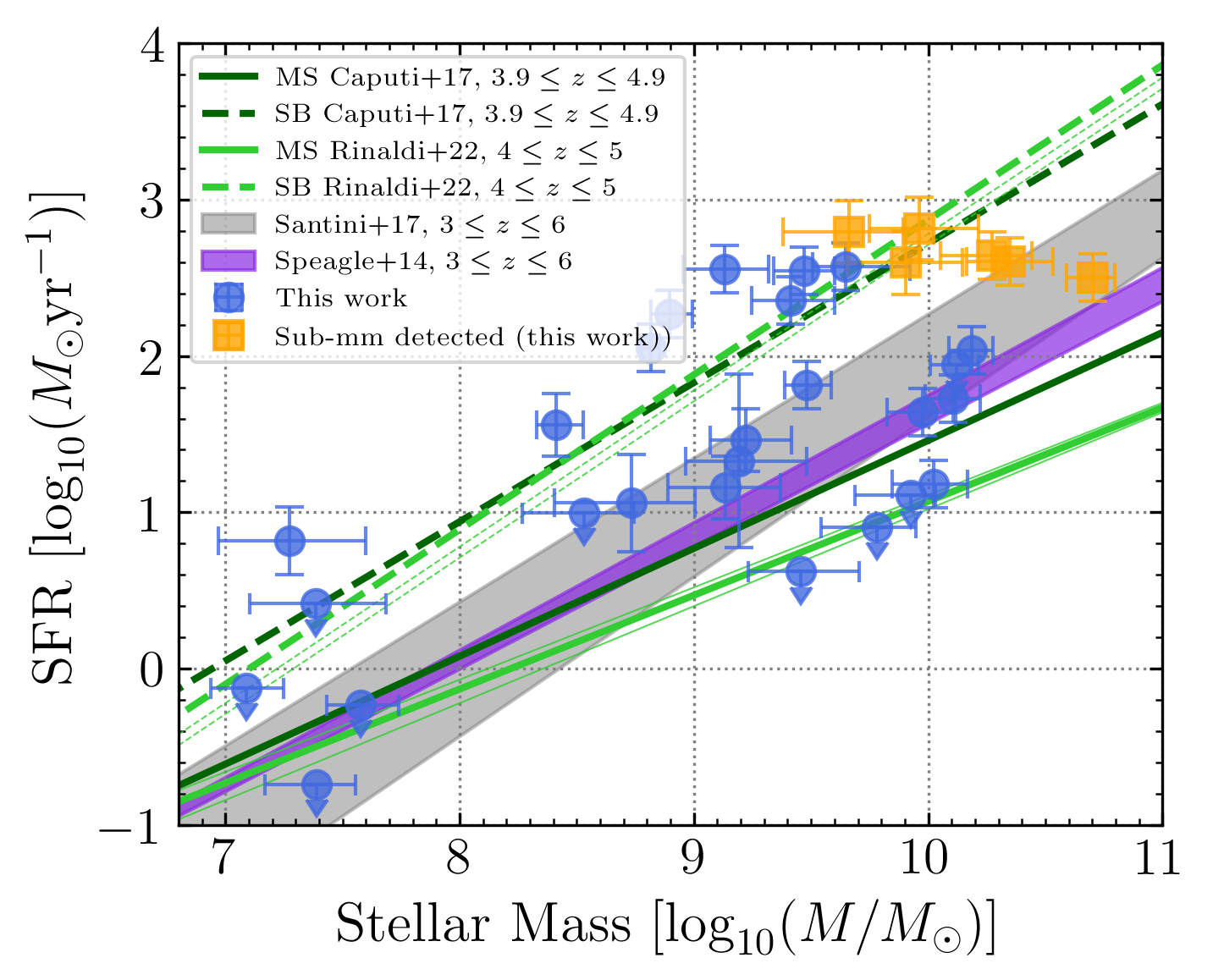}
    \caption{SFR derived from the H$\alpha$ luminosity vs. stellar mass for all JWST-detected galaxies in our sample. SFR upper limits are indicated with downward-pointing arrows. The orange data points represent the six submillimeter-detected galaxies in the sample. The solid and dotted lines display the MS and SB SFR-$M_*$ relations, respectively, taken from \citet{caputi2017} in dark green and \citet{rinaldi2022} in light green, respectively. The faint, light green lines correspond to the SFR-$M_*$ relations at $z=2.8$--4 and $z=5$--6 from \citet{rinaldi2022}. The MS relations between $z=3$ and 6 from \citet{speagle2014} and \citet{santini2017} are shown in violet- and gray-shaded regions respectively. \label{fig:halpha}}
\end{figure}

Finally, we show the H$\alpha$ emitters on the SFR-$M_*$ plane in Fig.\,\ref{fig:halpha}. We also indicate the main-sequence (MS) and starburst (SB) SFR-$M*$ relations from \citet{caputi2017} and \citet{rinaldi2022}, the former derived from H$\alpha$ emitters at $3.9 \leq z \leq 4.9$, the latter from star-forming galaxies in COSMOS at $2.8 \leq z \leq 6$. \highlight{We also show the MS relation for galaxies at $z=3$--6 from \citet{speagle2014} and \citet{santini2017}, although neither of these works separate SB and MS galaxies as done in \citet{caputi2017} and \citet{rinaldi2022}}. The H$\alpha$ SFRs span values between $\sim 5$--400 $M_\odot\,\mathrm{yr^{-1}}$, and as expected, broadly increase with the stellar mass. We find that our H$\alpha$ emitters are scattered over the SFR-$M*$ plane: out of the \highlight{23 JWST-detected galaxies with secure H$\alpha$ SFRs (that is, excluding upper-limit estimates and the aforementioned poorly constrained source PD-H-a-1/2 c), 8 are SBs with $\mathrm{sSFR} \geq -7.1$, and 9 are MS galaxies with $\mathrm{sSFR} \leq -8.1$. The remaining six sources have sSFRs between -8.0 and -7.7,} which following the classification from \citet{caputi2017} places them in the so-called star formation valley, such that they are likely transitioning from SBs into MS galaxies. \highlight{We point out that other works exploring the star formation valley at high redshift (e.g., \citealt{caputi2017,rinaldi2022}) typically describe this region as an underdensity. In our work instead, these galaxies constitute 25\% of the H$\alpha$ emitters, although it is difficult to derive any conclusions from this given the very specific source selection method employed in this paper, compared to the blind source analysis conducted in \citet{caputi2017} and \citet{rinaldi2022}}. 

\subsubsection{X-ray and submillimeter counterparts to the JWST-detected galaxies} \label{sect:submm}
Given the highly dusty nature of many of our JWST-detected galaxies, we search for counterparts in the literature, specifically at X-ray and submillimeter wavelengths. We find that source PD-H-s-2 has a $0\farcs7$ offset X-ray association from both the Chandra-COSMOS Legacy Survey \citep{civano2016} and the XMM-Newton Wide-Field Survey \citep{cappelluti2009}. Interestingly, as shown in Fig.\,\ref{fig:sed_secure}, the optical and near-infrared continuum of PD-H-s-2 is well fitted by stellar population templates, with $\chi^2_{\nu}=0.96$ and no preference from \textsc{LePhare} for an AGN solution. Compared to the best-fit SED, the photometry of PD-H-s-2 shows flux excess in the F200W and F356W bands, which sample the (H$\beta$+O[III]) and (H$\alpha$+ N[II] + S[II]) lines respectively. Finally, we notice that PD-H-s-2 and PD-H-s-1 have very similar SED properties and brightness (F444W $\sim$ 23.4 mag), but PD-H-s-1 lacks any ancillary detections at X-ray or radio wavelengths nor does it have similarly broad emission lines. 

\begin{figure}
    \gridline{\fig{seds/plot_SED_73050.png}{0.4\textwidth}{}}
    \gridline{\fig{seds/plot_SED_104114.png}{0.4\textwidth}{}}
    \caption{Best-fit SED obtained with \textsc{LePhare} on HST and JWST photometry for PD-H-s-2 and PD-Ks-d-10. Photometric measurements are shown in blue diamonds, flux upper limits in cyan with downward-pointing arrows, and template fluxes (with emission line contributions) with red circles. \label{fig:sed_secure}}
\end{figure}

\highlight{At the submillimeter regime, we find that sources PD-Ks-s-6, PD-Ks-s-8, PD-Ks-s-11, PD-H-d-1, PD-H-d-4, and PD-Ks-d-10 (all of which are H$\alpha$ emitters as shown in the previous section) are detected at submillimeter wavelengths with the Atacama Large Millimeter/submillimiter Array (ALMA) from the A$^3$COSMOS catalog \citep{liu2019}, with $\sim 1$--2 mJy fluxes from Band 6 ($\sim 1.2$ mm). In addition, PD-Ks-s-6, PD-Ks-s-8, and PD-H-d-10 were observed with SCUBA-2 at 850 $\mu$m and followed up at 870 $\mu$m with ALMA from the S2COSMOS study \citep{simpson2019} and the AS2COSMOS pilot study, respectively \citep{simpson2020}. PD-Ks-s-8 is also detected at 450 and 850 $\mu$m from the SCUBA-2 Cosmology Legacy Survey \citep{koprowski2016}. Finally, PD-H-d-1 is detected with the Very Large Array (VLA) at 3 GHz \citep{smolcic2017}.}

\citet{liu2019} report photometric redshifts of $z=5.9$ and $z=6.4$ and stellar masses of $\log(M_*/M_\odot) \sim 11.2$ for sources PD-Ks-s-6 and PD-H-d-1, respectively. These parameters are discrepant with our own solutions for these galaxies, although the probability distribution function of PD-Ks-s-6's redshift, PDF($z$), is degenerate with a secondary solution at $z\sim 6.7$. Even when we rerun \textsc{LePhare} on our photometry with the redshift fixed to $z=5.9$, we only retrieve a stellar mass of $\log(M_*/M_\odot) = 10.4$ for this source. \highlight{The \citet{liu2019} photometric redshifts for these two sources were actually taken from the superdeblended COSMOS catalog from \citet{jin2018}, who derived these redshifts from SED fitting far-infrared to submillimeter data. We therefore credit the discrepancy in redshifts to the different methodologies employed and different SED regimes considered for the SED fitting between their work and ours.} 

\highlight{We show the SED for PD-Ks-d-10 in Fig.\,\ref{fig:sed_secure}. With its photometric redshift $z=6.2$ and stellar mass of $\log(M_*/M_\odot)=10.7$, it is amongst the highest redshift, most massive galaxies in our dropout sample.} We have already shown the best-fit SED of PD-Ks-s-6 in Fig.\,\ref{fig:sed_halpha}. This source is amongst the four galaxies in our sample that would qualify as a true $K_{\mathrm{s}}$-band dropouts, given that it has flux detections of at most $\sim 1\sigma$ significance in any of the bands redward of the F200W band, nor is it visibly by eye in said bands. Although its nature is not particularly different from the other $z\sim4.7$ galaxies in our sample, PD-Ks-s-6 is highly dusty with $E(B-V)=0.7$ and the third reddest source in the sample with [F160W-F444W]=4.1. 

\section{Discussion and conclusions} \label{sect:conclusion}

In this work, we have studied a sample of Spitzer IRAC-detected galaxies that are part of the SMUVS program in the COSMOS field but which are completely undetected in the latest data release (DR5) of the UltraVISTA VIRCAM $H$- and $K_{\rm s}$-band imaging. These sources are rare, as we identify only \highlight{26} over an area of 94 arcmin$^2$ (i.e., the area with multiwavelength JWST coverage considered in this work). We used the available JWST/NIRCam data in seven bands from the PRIMER program, together with ancillary HST/ACS and HST/WFC3 data, to study the properties of these galaxies,  with vastly improved angular resolution and sensitivity with respect to the original Spitzer and ground-based data. We find that \highlight{more than half} of the IRAC sources are in fact systems of multiple galaxies: 10 consisting of a so-called deboosted galaxy that is accompanied by a low-redshift, faint-foreground contaminant; and another 5 consisting of associated galaxies at similar redshifts. In addition, \highlight{the majority of} the dropout candidates display flux in the short-wavelength JWST channels, such that only \highlight{two $K_{\rm s}$-band dropout candidates and two $H_{\rm s}$-band dropout candidates} qualify as true dropout galaxies.

We find that our JWST-detected galaxies display a variety of properties, with redshifts between $z=0.8$--6.2, stellar masses $\log(M_*/M_\odot)=7.0$--10.7, and color excess between $E(B-V)=0.0$ and 1.0. Most of the sample consists of intermediate-mass galaxies that are red because of dust attenuation, with median values of \highlight{$z=3.6$, $\log(M_*/M_\odot)=9.5$, [F160W-F444W]$=2.2$, and $E(B-V)=0.5$.} We compared the properties of our JWST-detected galaxies at $z=3$--6 with those of UltraVISTA DR4 $HK_{\rm s}$-detected galaxies in the same region. We find that our JWST-detected galaxies are amongst the dustiest galaxies at this epoch; in fact, \highlight{$\sim 65$\,\%} of our sample at $z=3$--6 has $E(B-V)\geq0.5$, compared to 7\% at most amongst the UltraVISTA DR4 galaxies. 

For \highlight{$\sim 75$\%} of our IRAC-selected dropout candidates, the JWST photometry shows flux excess in bands encompassing the H$\alpha$ emission-line complex, which enabled us to calculate (H$\alpha$+ N[II] + S[II]) rest-frame equivalent widths and H$\alpha$ star formation rates. The (H$\alpha$+ N[II] + S[II]) $\mathrm{EW_{0}}$ span values between \highlight{$\sim 100$ and 2200\,\AA}, and the SFRs between \highlight{$\sim 5$ and 375} $M_\odot\, \mathrm{yr^{-1}}$. We studied them in the SFR-$M_*$ plane and found \highlight{35\,\%} of the JWST-detected galaxies \highlight{with secure H$\alpha$ SFRs} to be SBs, \highlight{40\,\%} to be MS galaxies and the remaining 25\,\% to be located in the star-formation valley.  Given their redshifts, the H$\alpha$ emission contributes strongly to the continuum measurements redward of the $K_{\rm s}$ band, explaining, together with their dusty nature, why these galaxies enter our initial dropout sample. 

We checked if the H$\alpha$-emitting nature of our sample is apparent from the IRAC photometry as well. For seven dropout candidates, the H$\alpha$ line falls in the 3.6 $\mu$m band, such that we would expect to see a flux excess with respect to the 4.5 $\mu$m band. However, for only \highlight{three} of these galaxies do we observe $\mathrm{m_{3.6}-m_{4.5}} < 0$, such that we conclude the presence of H$\alpha$ emission is not obvious from the IRAC bands alone. We note that although there are multiple dropout candidates for which the H$\alpha$ line is observed with the 4.5$\mu$m band, we cannot perform a similar check, as a flux excess in this band with respect to the $3.6 \mu$m band can be explained by red continuum emission just as well.

We have crossmatched our sources with ancillary data at X-ray and submillimeter wavelengths, and find one AGN candidate and \highlight{six} ALMA-detected galaxies amongst them. For the ALMA-detected sources we obtained photometric redshifts between $z\sim3.3$ and 6.2 and stellar masses $\log(M_*/M_\odot)=9.7$--10.7 from the HST and JWST photometry. These sources are among the (H$\alpha$+ N[II] + S[II]) line emitters discussed above. 

Our galaxies are not as Spitzer/IRAC-bright and massive as other studies targeting optically faint sources. The brightest IRAC source in our sample is 23.9 and 23.1 mag in the $3.6 \mu$m and $4.5\mu$m bands, respectively, and the median IRAC magnitude of our dropout candidates is $\sim 24.6$ mag. In addition, the most massive galaxy in our sample is $\log(M_*/M_\odot)=10.7$ and the median stellar mass of the sample is $\log(M_*/M_\odot)=9.5$. Similar endeavors to ours have been focused mostly on bright IRAC sources ($[4.5] < 24$ mag), finding them to be massive $\log(M_*/M_\odot)>10.5$ star-forming galaxies at $z\sim3$--6. \citet{alcalde2019} do find H$\alpha$ emitters among relatively Spitzer-bright ($[4.5]<24.5$ mag) HST/F160W-band dropout sources, although their incidence is only a third in their sample, versus our finding of H$\alpha$ emitters among \highlight{$\sim75$ \,\%} of our dropout candidates. More recently, \citet{barrufet2023} looked for JWST-detected HST-dropout galaxies in CEERS. Their sample spans a similar NIRCam/F444W magnitude range as our JWST-detected galaxies (23--28 mag), which they find to be dusty $E(B-V)\sim0.5$, relatively massive $\log(M_*/M_\odot)\sim 10$ star-forming galaxies at $z\sim2$--8. The presence of H$\alpha$ line emission is unfortunately not discussed in their work. 

Lastly, our dropout candidates, with an observed surface density of \highlight{$\sim 0.3$ arcmin$^{-2}$, appear to be relatively rare}. Their incidence is comparable to that of mid-infrared bright, true HST-dropout galaxies, for which studies have found typical surface densities of $\sim 0.1$ arcmin$^{-2}$ (e.g., \citealt{alcalde2019,wang2019}). However, a recent study by \citet{barrufet2023} reported a surface density of $\sim 0.8$ arcmin$^{-2}$ for JWST-detected, HST-dark galaxies.

In summary, we have revealed the nature of a representative sample of the IRAC sources that were left unidentified from the Spitzer era. In the future, once JWST/NIRCam imaging becomes available for other well-studied extragalactic fields in the sky, we will be able to follow up more of these sources and \highlight{characterize the overall population of emission-line galaxies} with significant dust extinction at intermediate/high redshifts. 

\begin{acknowledgments}
 We thank the anonymous referee for a constructive report. We thank Ian Smail for updating our information on the presence of sub-millimetre sources within our galaxy sample, and Matt Jarvis for their useful feedback. KIC and VK acknowledge funding from the Dutch Research Council (NWO) through the award of the Vici Grant VI.C.212.036. This work is based on observations made with the NASA/ESA/CSA James Webb Space Telescope. The data were obtained from the Mikulski Archive for Space Telescopes at the Space Telescope Science Institute, which is operated by the Association of Universities for Research in Astronomy, Inc., under
NASA contract NAS 5-03127 for JWST. All the JWST data used in this paper can be found in MAST: \href{https://archive.stsci.edu/doi/resolve/resolve.html?doi=10.17909/bysp-ds64}{10.17909/bysp-ds64}.
\end{acknowledgments}

\appendix
\section{HST and JWST image cutouts of 26 dropout candidates} \label{appendix:image_cutouts}
\begin{figure*}[h!]
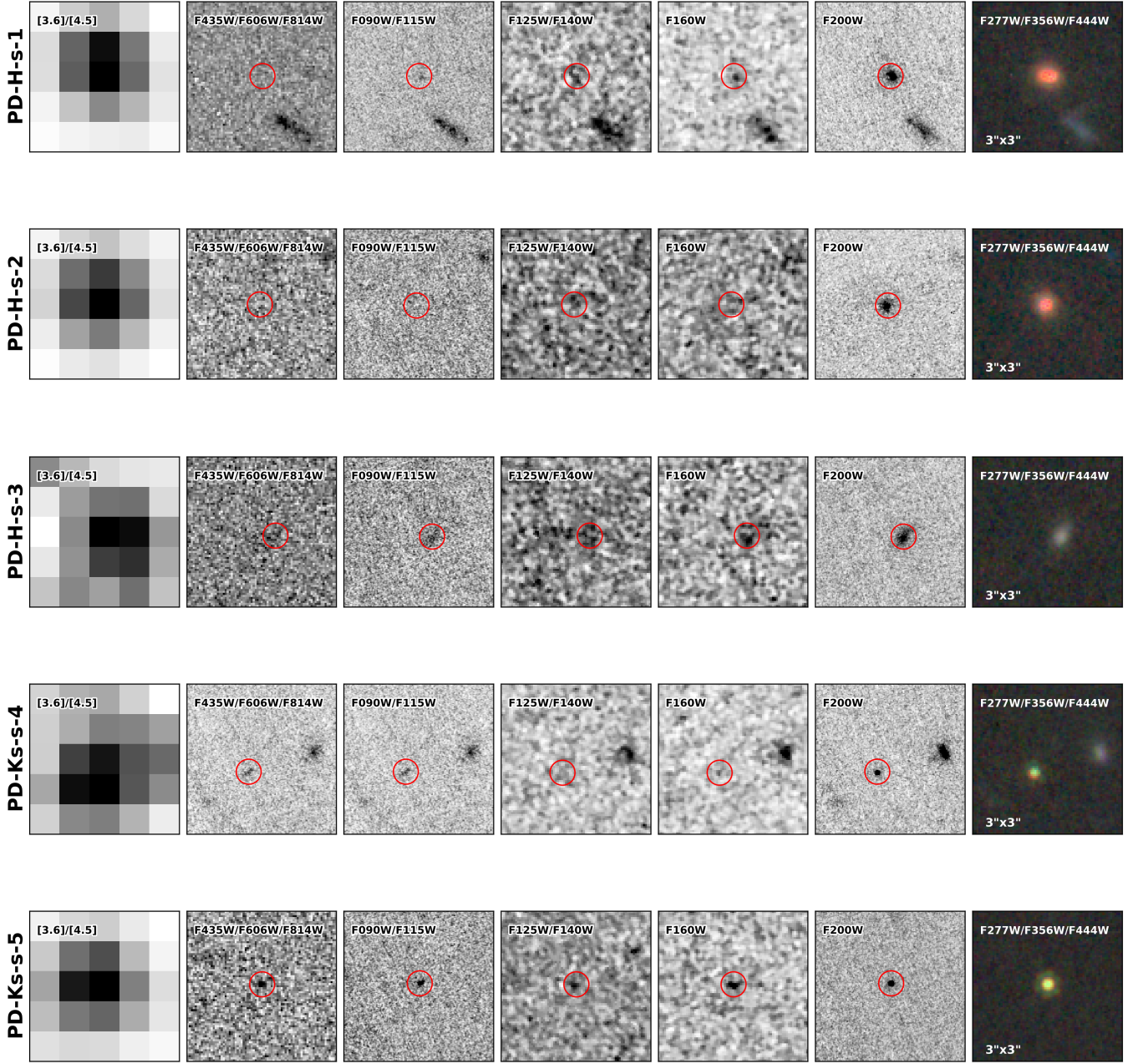

\gridline{\fig{stamps_paper_PD-H-s-1.png}{\textwidth}{}}
\gridline{\fig{stamps_paper_PD-H-s-2.png}{\textwidth}{}}
\gridline{\fig{stamps_paper_PD-H-s-3.png}{\textwidth}{}}
\gridline{\fig{stamps_paper_PD-Ks-s-4.png}{\textwidth}{}}
\gridline{\fig{stamps_paper_PD-Ks-s-5.png}{\textwidth}{}}
\caption{Image cutouts for the 11 secure dropout candidates. The red circle represents the $0\farcs5$ diameter circular aperture at the position of the JWST-detected galaxy. 
\label{fig:stamps_secure_1}}
\end{figure*}

\begin{figure*}[h!]
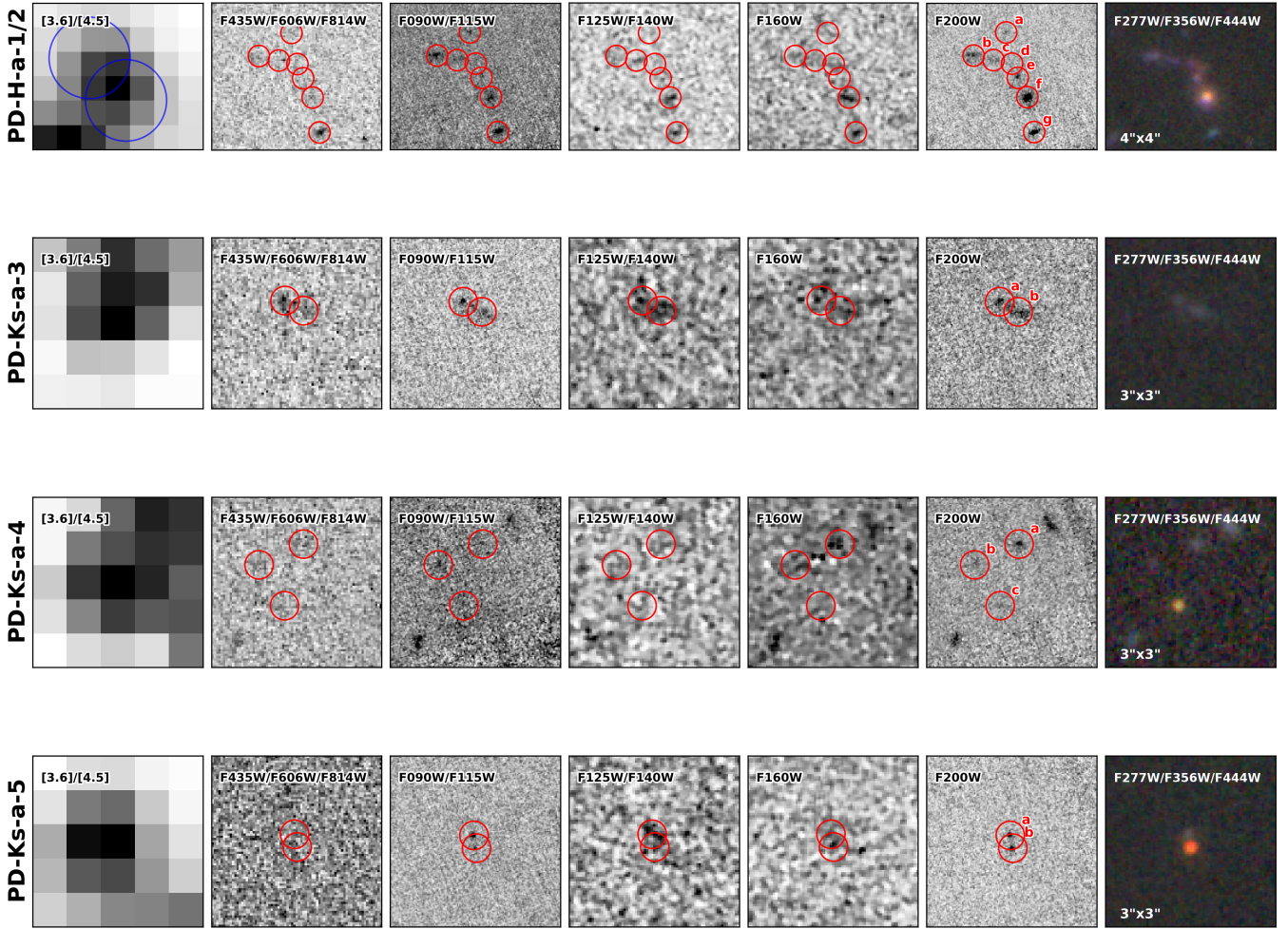

\gridline{\fig{stamps_paper_PD-H-a-1-2forpaper.png}{\textwidth}{}}
\gridline{\fig{stamps_paper_PD-Ks-a-3_forpaper.png}{\textwidth}{}}
\gridline{\fig{stamps_paper_PD-Ks-a-4_forpaper.png}{\textwidth}{}}
\gridline{\fig{stamps_paper_PD-Ks-a-5_forpaper.png}{\textwidth}{}}
\caption{Image cutouts for the 5 associated dropout candidates. The red circles represent the $0\farcs5$ diameter circular aperture at the positions of the JWST-detected galaxies, which are annotated with their IDs on the F200W image cutout. As PD-H-a-1/2 is a system consisting of two IRAC detections, we indicate the $2\arcsec$ IRAC apertures on the 3.6 and 4.5 $\mu$m image stack in blue. 
\label{fig:stamps_associated}}
\end{figure*}

In this appendix, we show the image cutouts for the 26 dropout candidates identified in this work. For each dropout candidate, from left to right, the horizontal panels show the following imaging in 3 by 3$"$ windows: IRAC 3.6 and 4.5 $\mu$m average; HST/ACS F435W, F606W, and F814W average; JWST/NIRCam F090W and F115W average; HST/WFC3 F125W and F140W average; JWST/NIRCam F200W; JWST/NIRCam F277W, F356W, and F444W rgb image. Figures \ref{fig:stamps_secure_1} and \ref{fig:stamps_secure_2} show the image cutouts for the 11 secure dropout candidates; Figures \ref{fig:stamps_deboosted_1} and \ref{fig:stamps_deboosted_2} show the image cutous for the 10 deboosted dropout candidates; Fig.\,\ref{fig:stamps_associated} shows the images cutouts for the five associated dropout candidates. 

\begin{figure*}[h!] 
\gridline{\fig{stamps_paper_PD-Ks-s-6.png}{\textwidth}{}}
\gridline{\fig{stamps_paper_PD-Ks-s-7.png}{\textwidth}{}}
\gridline{\fig{stamps_paper_PD-Ks-s-8.png}{\textwidth}{}}
\gridline{\fig{stamps_paper_PD-Ks-s-9.png}{\textwidth}{}}
\gridline{\fig{stamps_paper_PD-Ks-s-10.png}{\textwidth}{}}
\gridline{\fig{stamps_paper_PD-Ks-s-11.png}{\textwidth}{}}
\caption{Figure \ref{fig:stamps_secure_1} continued. 
\label{fig:stamps_secure_2}}
\end{figure*}

\begin{figure*}[h!]
\gridline{\fig{postage_stamps_deboosted/stamps_paper_PD-H-d-1.png}{\textwidth}{}}
\gridline{\fig{postage_stamps_deboosted/stamps_paper_PD-H-d-2.png}{\textwidth}{}}
\gridline{\fig{postage_stamps_deboosted/stamps_paper_PD-H-d-3.png}{\textwidth}{}}
\gridline{\fig{postage_stamps_deboosted/stamps_paper_PD-H-d-4.png}{\textwidth}{}}
\gridline{\fig{postage_stamps_deboosted/stamps_paper_PD-Ks-d-5.png}{\textwidth}{}}
\gridline{\fig{postage_stamps_deboosted/stamps_paper_PD-Ks-d-6.png}{\textwidth}{}}
\gridline{\fig{postage_stamps_deboosted/stamps_paper_PD-Ks-d-7.png}{\textwidth}{}}
\caption{Image cutouts for the 10 deboosted dropout candidates. The circles represent the $0\farcs5$ diameter circular aperture at the positions of the JWST-detected galaxies, showing the deboosted source in red and the contaminating foreground galaxies in blue. 
\label{fig:stamps_deboosted_1}}
\end{figure*}

\begin{figure*}[h!]

\gridline{\fig{stamps_paper_PD-Ks-d-8.png}{\textwidth}{}}
\gridline{\fig{stamps_paper_PD-Ks-d-9.png}{\textwidth}{}}
\gridline{\fig{stamps_paper_PD-Ks-d-10.png}{\textwidth}{}}
\caption{Figure \ref{fig:stamps_deboosted_1} continued.
\label{fig:stamps_deboosted_2}}
\end{figure*}

\end{document}